\documentclass[twocolumn,american,english,aps,prb,showpacs,amsmath,amssymb,superscriptaddress,a4]{revtex4}
\usepackage[active]{srcltx}
\usepackage{color}
\usepackage{array}
\usepackage{textcomp}
\usepackage{multirow}
\usepackage{amsmath}
\usepackage{amssymb}
\usepackage{wasysym}
\usepackage{graphicx}
\usepackage{esint}
\usepackage{subscript}

\makeatletter

\newcommand{\noun}[1]{\textsc{#1}}
\DeclareRobustCommand{\greektext}{%
  \fontencoding{LGR}\selectfont\def\encodingdefault{LGR}}
\DeclareRobustCommand{\textgreek}[1]{\leavevmode{\greektext #1}}
\DeclareFontEncoding{LGR}{}{}
\DeclareTextSymbol{\~}{LGR}{126}
\newcommand{\lyxmathsym}[1]{\ifmmode\begingroup\def\b@ld{bold}
  \text{\ifx\math@version\b@ld\bfseries\fi#1}\endgroup\else#1\fi}

\providecommand{\tabularnewline}{\\}

\@ifundefined{textcolor}{}
{%
 \definecolor{BLACK}{gray}{0}
 \definecolor{WHITE}{gray}{1}
 \definecolor{RED}{rgb}{1,0,0}
 \definecolor{GREEN}{rgb}{0,1,0}
 \definecolor{BLUE}{rgb}{0,0,1}
 \definecolor{CYAN}{cmyk}{1,0,0,0}
 \definecolor{MAGENTA}{cmyk}{0,1,0,0}
 \definecolor{YELLOW}{cmyk}{0,0,1,0}
}

\usepackage{multirow}\usepackage{bm}\usepackage{gensymb}\usepackage{threeparttable}

\makeatother

\usepackage{babel}
\begin{document}

\title{\textcolor{black}{The importance of anisotropic Coulomb interaction
in LaMnO\textsubscript{%
3%
}}}

\author{Thomas A. Mellan}

\affiliation{Department of Chemistry, University College London, 20 Gordon Street,
London WC1H 0AJ, United Kingdom}

\author{Furio Cora}

\affiliation{Department of Chemistry, University College London, 20 Gordon Street,
London WC1H 0AJ, United Kingdom}

\author{Ricardo Grau-Crespo}

\affiliation{Department of Chemistry, University of Reading, Whiteknights, Reading
RG6 6AD, United Kingdom}

\author{Sohrab Ismail-Beigi}

\affiliation{Department of Applied Physics, Yale University, New Haven, Connecticut,
USA}

\affiliation{Center for Research on Interface Structures and Phenomena (CRISP),
Yale University, New Haven, Connecticut, USA}

\begin{abstract}
In low-temperature anti-ferromagnetic LaMnO$_{3}$, strong and localized
electronic interactions among Mn 3\emph{d} electrons prevent a satisfactory
description from standard local density and generalized gradient approximations
in density functional theory calculations. Here we show that the strong
on-site electronic interactions are described well only by using direct
\emph{and} exchange corrections to the intra-orbital Coulomb potential.
Only DFT+\emph{U} calculations with explicit exchange corrections
produce a balanced picture of electronic, magnetic and structural
observables in agreement with experiment. To understand the reason,
a rewriting of the functional form of the +\emph{U} corrections is
presented that leads to a more physical and transparent understanding
of the effect of these correction terms. The approach highlights the
importance of Hund's coupling (intra-orbital exchange) in providing
anisotropy across the occupation and energy eigenvalues of the Mn
\emph{d} states. This intra-orbital exchange is the key to fully activating
the Jahn-Teller distortion, reproducing the experimental band gap
and stabilizing the correct magnetic ground state in LaMnO\textsubscript{%
3%
}. The best parameter values for LaMnO\textsubscript{3} within the
DFT (PBEsol) +\emph{U} framework are determined to be $U=8$ eV and
$J=1.9$ eV.
\end{abstract}


\maketitle

\section{Introduction}

LaMnO$_{3}$ (LMO) is characteristic of the ABO\textsubscript{%
3%
} family of strongly correlated transition metal oxide perovskites,
which generally exhibit complex phase diagrams, as a result of subtle
coupling across several distinct mechanisms.\citep{Franchini} Bulk,
thin film, and interfacial LaMnO$_{3}$ are subject to a multitude
of symmetry breaking mechanisms, including crystal field,\citep{Franchini}
octahedral distortion,\citep{Solovyev1996} orbital ordering and Jahn-Teller
distortion,\citep{Feiner1999,Millis1997,Millis1998,Baldini2011} Mott-type
strong \emph{d} electron Coulomb interactions (direct and exchange),\citep{Loa2001,Tokura2000}
and charge transfer ordered (Verwey) states\citep{M.UeharaS.MoriC.H.Chen1999,Chu2013,Coey2004}.
All of these mechanisms are believed to exist and compete in varying
ways in this material. As a result, LaMnO$_{3}$ naturally exhibits
a rich phase diagram as a function of temperature and pressure\citep{Baldini2011}
as well as dopant concentration,\citep{Tokura2006,Chen2012} which
together make LaMnO$_{3}$ the single most examined metal oxide in
the LaXO$_{3}$ class (where X is a transition metal atom).\citep{He2012}
Doping on the ABO\textsubscript{%
3%
} A site provides a particularly rich field of experimentally observed
phenomena, with both Na and Ca doped La$_{1-x}$A$_{x}$MnO$_{3}$
exhibiting colossal magneto resistance (CMR)\citep{Search,N.D.M1973}
and a Seebeck coefficient that can exhibit positive or negative values
which may lead to potential thermopower applications.\citep{Mahendiran1996}
Pure bulk LaMnO\textsubscript{%
3%
} is spin polarized and non-polar, but recent theoretical work shows
that the magnetic state in Sr doped La$_{1-x}$Sr$_{x}$MnO$_{3}$
may be controlled through variation in the electric polarization state.\citep{Chen2012}
Recent multi-ferroic theory predicts novel magnetic properties due
to \emph{t}\textsubscript{%
2%
}\emph{\textsubscript{%
\emph{g}%
}} ferromagnetic superexchange in Ti doped LMO interfaces.\citep{Garcia-Barriocanal2010}
Finally, the interface between La$_{1-x}$Sr$_{x}$MnO$_{3}$ and
a ferroelectric shows a polar state that also has a reversible orbital
polarization.\textcolor{black}{\citep{Chen2014a}}

This interest in LMO from condensed matter and materials scientists
underscores the value of a reliable first principles description based
on, for example, density functional theory (DFT). In particular the
magnetic, electronic and crystal structure should be accessible simultaneously
within a low-cost computational framework. Unfortunately previous
Hartree-Fock, DFT and hybrid-functional examinations of bulk LMO show
that that obtaining a satisfactory description is not trivial.\citep{Munoz2004,Franchini} 

In this work we show the limitations and successes of two different
DFT+\emph{U} methods. The Dudarev \textit{et al}. Coulomb correction,\citep{Dudarev1998}
here called $U_{\text{eff}}$, averages out exchange effects of the
Mn \textit{d} shell, and we show that it cannot simultaneously reproduce
the bulk band gap, structure and magnetism. The dedicated anisotropic
exchange term in the Liechtenstein \textit{et al.} Coulomb correction,\citep{Liechtenstein1995a}
here called $U|J$, dramatically improves the description of LMO.
The $U|J$ method answers the specific call for a practical DFT-based
methodology capable of reproducing the gap, structure and magnetism
simultaneously in LMO.\citep{Sawada1997} This is useful as understanding
the coupling between electronic, magnetic and lattice degrees of freedom
in LMO is a matter of persistent interest.\citep{Sawada1997,Solovyev1996,Hashimoto2010} 

Using the $U|J$ method we show the importance of \textcolor{black}{Hund's
coupling} in LMO. Intra-orbital exchange can energetically order the
orbitals of the Mn \emph{t}\textsubscript{%
2\emph{g}%
}\textsuperscript{%
3%
}\emph{e\textsubscript{%
g%
}}\textsuperscript{%
1%
} ion, which in turn strongly affects \emph{inter}-\emph{orbital} magnetism
and the size of the LMO band gap. The Mn \emph{e\textsubscript{%
g%
}}\textsuperscript{%
1%
} occupancy polarization\citep{Wu2013,Chen2014a}

\begin{equation}
\pi^{e_{g}\sigma}=\frac{f_{x^{2}-y^{2}\sigma}-f_{3z^{2}-r^{2}\sigma}}{f_{x^{2}-y^{2}\sigma}+f_{3z^{2}-r^{2}\sigma}}\:,\label{eq:orbital_polarisation_definition}
\end{equation}
for the $x^{2}-y^{2}$ and $3z^{2}-r^{2}$ occupancy eigenvalues $(f)$
where $\sigma$ labels spin, is highly sensitive to intra-orbital
exchange term \emph{$J$} in the $U|J$ scheme. By modifying the sign
and value of $\pi^{e_{g}\sigma}$, we correct the DFT description
of Jahn-Teller (JT) distortion, and the electronic and magnetic structures
of LMO. In addition the $U|J$ calculations provide insight into the
origin of magnetic, electronic and structural ordering in LMO.

\section{Methodology}

Periodic plane wave density functional theory (DFT) calculations are
performed using the VASP software,\citep{Kresse1996,Kresse1996a}
the local density approximation (LDA PZ81)\citet{Perdew1981}, and
the generalized gradient approximation (GGA) in the form of the Perdew-Burke-Erzenhof
solids-adapted exchange correlation functional (PBEsol).\citep{Perdew2008,Perdew1996}
Valence electrons are described using the projector augmented wave
(PAW) method\citep{Blochl1994,Kresse1999} with core states (up to
$4$\emph{d} in La, $2$\emph{p} in Mn, and ${\color{black}2}$\emph{s}
in O) frozen at their atomic reference states. Plane-waves were cutoff
above a kinetic energy of $520$ eV, and a $5\times4\times5$ \emph{$\mathbf{k}$}-point
mesh of was employed for the LaMnO$_{3}$ unit cells. All relaxed
structures fulfill a \textcolor{black}{convergence criterion of less
than $0.01$ eV/${\rm \AA}$, for both ionic forces and volume-normalized
stresses (as standard in VASP).}\textcolor{red}{{} }

DFT has known shortcomings in the prediction of the electronic structure
of materials with localized electronic states.\citep{anisimov2010electronic,Anisimov1997,Liechtenstein1995a}
A typical example are the bands derived from Mn \emph{d} orbitals
in LaMnO\textsubscript{%
3%
}: the errors can be corrected to various extents by employing Hubbard-U
type corrections to account for intra-atomic Coulomb interactions
in the DFT+\emph{U} approach.\citep{Anisimov1997,Dudarev1998,Liechtenstein1995a}
The most popular and simplest Coulomb correction is the ``Spherically
Averaged'' scheme of Dudarev \emph{et al.},\citep{Dudarev1998} here
called DFT+$U_{\text{eff}}$, which has only a single effective \emph{U}
parameter, $U_{\text{eff}}$. A more sophisticated approach is the
``Rotationally Invariant'' scheme of Lichtenstein and Zaanen,\citep{Liechtenstein1995a}
which we label here as DFT+$U|J$. Note the simpler Dudarev $U_{\text{eff}}$
approach was developed after the Liechtenstein $U|J$ approach, and
both are fully rotationally invariant. 

Both DFT+\emph{U} methodologies add Hartree-Fock type corrections
to the DFT total energy that act on a local sub-space of atomic-like
orbitals. The DFT+$U_{\text{eff}}$ total energy is given by

\begin{equation}
E_{\text{DFT}+U_{\text{eff}}}=E_{\text{DFT}}+\frac{U_{\text{eff}}}{2}\sum_{at}\sum_{i,\sigma}(f_{i\sigma}-f_{i\sigma}^{2})\label{eq:dudarevenergy}
\end{equation}
where $E_{\text{DFT}}$ refers to some chosen flavor of electron density-based
exchange-correlation approximation (LDA or GGA in our work). The index
$at$ specifies the Mn sites where the correction is performed. The
eigen-occupations $f_{i\sigma}$ of the electronic on-site density
matrix are labeled by \textcolor{black}{spin $\sigma$ and index }\textit{\textcolor{black}{i
}}\textcolor{black}{which represents a linear combination of angular
momentum quantum numbers (which in our case ranges over the five magnetic
quantum numbers} \textcolor{black}{$m=-2,-1,0,1,2$ for the $3$}\emph{d}\textcolor{black}{{}
Mn orbitals)}. $U_{\text{eff}}=U-J$ is the Hubbard-type energy parameter
for this approach while $U$ and \emph{J} are the separate direct
and exchange Coulomb parameters\citep{Liechtenstein1995a} (see also
Appendix A). 

For our work here, the DFT+$U|J$ total energy is best rewritten as
an added correction to the DFT+$U_{\text{eff}}$ approach (as detailed
in Appendix A) given by
\begin{multline}
E_{\text{DFT}+U|J}=E_{\text{DFT}}+E_{\text{corr}}\\
=E_{\text{DFT}}+\frac{U_{\text{eff}}}{2}\sum_{at}\sum_{i,\sigma}(f_{i\sigma}-f_{i\sigma}^{2})\,+\\
\frac{1}{2}\sum_{\sigma\sigma',ij}\text{C}_{ij}^{\,\sigma\sigma'}f_{i\sigma}f_{j\sigma'}-\Delta\text{X}_{ij}^{\,\sigma}f_{i\sigma}f_{j\sigma}\delta_{\sigma\sigma'}\,.\label{eq:Lichtenstein RI energy}
\end{multline}
The correction to the DFT band energy eigenvalue $\epsilon_{i\sigma}$
stems from the occupancy derivative of the correction terms given
by
\begin{multline*}
\Delta\epsilon_{i\sigma}^{\text{corr}}=\frac{\partial E_{\text{corr}}}{\partial f_{i\sigma}}=U_{\text{eff}}\left(\frac{1}{2}-f_{i\sigma}\right)\\
+\sum_{j,\sigma'}\text{C}_{ij}^{\,\sigma\sigma'}f_{j\sigma'}-\Delta\text{X}_{ij}^{\,\sigma}f_{j\sigma}\delta_{\sigma\sigma'}\,,
\end{multline*}
where the first term is the $U_{\text{eff}}$ correction and the second
and third terms are the added contribution from the $U|J$ scheme.
For compactness and for use below, it is useful to collect all occupancies
or energy eigenvalues for the same spin into a vector $f_{\sigma}$
or $\epsilon_{\sigma}$ in order write these corrections in matrix
notation. For atomic \emph{d} shells, the Appendix A shows that 

\begin{equation}
\Delta\epsilon_{\sigma}^{\text{corr}}=U_{\text{eff}}\left(\frac{1}{2}-f_{\sigma}\right)+J\text{A}^{\sigma}f_{\sigma}+J\text{B}^{\sigma}f_{\bar{\sigma}}\label{eq:energyshiftRIUJformula}
\end{equation}
where $\bar{\sigma}$ represents the opposite spin to $\sigma$. For
canonical $t_{2g}$ and $e_{g}$ orbitals, the dimensionless matrices
$\text{A}^{\sigma}$ and $\text{B}^{\sigma}$ are
\[
\mathrm{A}^{\sigma}=\left(\begin{array}{r|rrrrr}
 & 3z^{2}-r^{2} & x^{2}-y^{2} & xy & yz & xz\\
\hline 3z^{2}-r^{2} & 0 & -0.52 & -0.52 & 0.52 & 0.52\\
x^{2}-y^{2} & -0.52 & 0 & 0.86 & -0.17 & -0.17\\
xy & -0.52 & 0.86 & 0 & -0.17 & -0.17\\
yz & 0.52 & -0.17 & -0.17 & 0 & -0.17\\
xz & 0.52 & -0.17 & -0.17 & -0.17 & 0
\end{array}\right)
\]
and
\[
\mathrm{B}^{\sigma}=\left(\begin{array}{r|rrrrr}
 & 3z^{2}-r^{2} & x^{2}-y^{2} & xy & yz & xz\\
\hline 3z^{2}-r^{2} & 1.14 & -0.63 & -0.63 & 0.06 & 0.06\\
x^{2}-y^{2} & -0.63 & 1.14 & 0.29 & -0.40 & -0.40\\
xy & -0.63 & 0.29 & 1.14 & -0.40 & -0.40\\
yz & 0.06 & -0.40 & -0.40 & 1.14 & -0.40\\
xz & 0.06 & -0.40 & -0.40 & -0.40 & 1.14
\end{array}\right)\,.
\]

Both DFT+\emph{U }methodologies permit the description of electron
localisation phenomena, that stem from Hartree-Fock physics and the
related removal of self-interaction errors, which enable essential
long-range ordering (orbital, spin, charge and lattice degrees of
freedom).\citep{anisimov2010electronic} For Mn in LMO, delocalised
\emph{s} and \emph{p} orbitals typify the weakly correlated electronic
states successfully described by DFT, while the localised Mn \emph{d}\textsuperscript{%
} states require the +\emph{U} correction. In the DFT+$U_{\text{eff}}$
approach,\emph{ }$U_{\text{eff}}$\emph{ }in Equation (\ref{eq:dudarevenergy})
provides occupation-dependent corrections to DFT, while the DFT+$U|J$
approach in Equation (\ref{eq:Lichtenstein RI energy}) adds further
degrees of explicit spatial/orbital dependent corrections. Both corrections
provide a basis for for energy splitting of \emph{d} orbitals (and
related symmetry breaking and orbital polarization) on top of splittings
due to spin exchange and/or crystalline geometrical distortions already
present at the LDA or GGA density functional level.

The $U|J$ correction variety in Equation (\ref{eq:energyshiftRIUJformula})
is most relevant to materials with strongly interacting electrons
with an explicit orbital symmetry dependence,\citep{Himmetoglu2014}
for example, Fe-based superconductors,\citep{Nakamura2009} heavy
fermion metals,\citep{Jeong2006} non-collinear magnetic materials,\citep{Bousquet2010,Tompsett2012}
and orbitally ordered materials in which Hund's coupling is critical
to establishing the correct insulating or metallic character.\citep{Himmetoglu2011}
Although the anisotropic exchange corrections to DFT have successfully
been used to describe manganese oxides in the past,\citep{Mellan2013a,Cockayne2012,Tompsett2012}
we believe our work is the first explicit calculation and analysis
of the $U|J$ exchange matrix elements and anisotropic splitting for
LaMnO\textsubscript{%
3%
}. 

\begin{figure}
\includegraphics[scale=0.18]{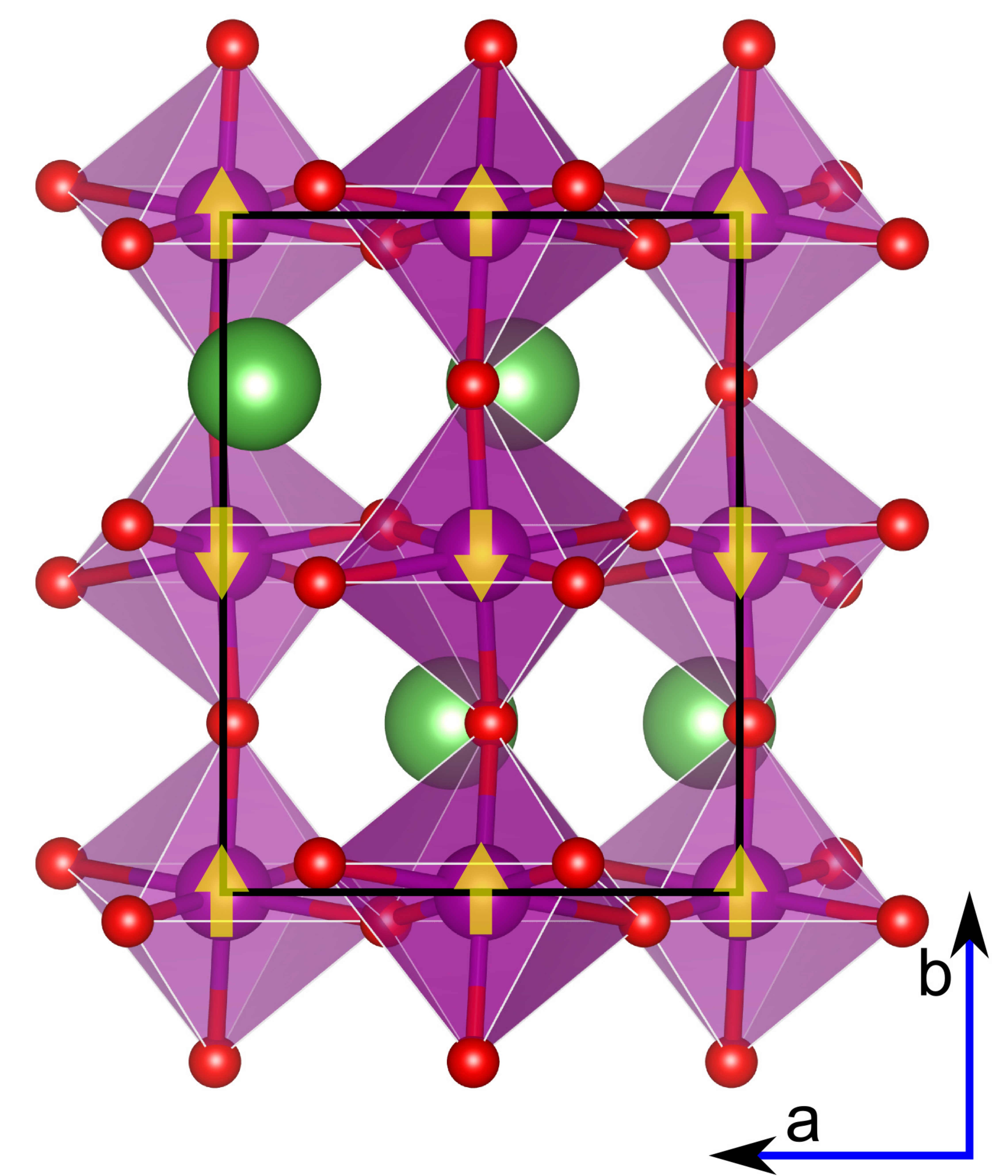}

\caption{%
$(001)$ face of A-type antiferromagnetic (A-AFM) LaMnO$_{3}$ . Mn
in purple,\emph{ }La\emph{ }in green and\emph{ }O\emph{ }in red. Arrows
indicate direction of spin polarization on Mn ions.%
}
\label{fig:exptlmostruct}
\end{figure}

\textcolor{black}{In our work, our global coordinate system is chosen
to align the orthogonal }\textcolor{black}{\emph{$x',y,'z'$}}\textcolor{black}{{}
axes along the L}MO\textcolor{black}{{} unit cell ($a,b,c$) vectors.
A local }\textcolor{black}{\emph{$x,y,z$}}\textcolor{black}{{} basis
for each Mn is defined by aligning the local axes with the Mn-O bonds
of each tilted MnO\textsubscript{%
6%
} octahedron (see Figures \ref{fig:exptlmostruct} and \ref{fig:Orbital-ordering-across-LMO}):
the local $x$ axis is chosen along the shortest Mn-O bond (strongly
JT active), the local $y$ axis along the intermediate length Mn-O
bond (here called apical), and the local $z$ axis is along the longest
Mn-O bond (strongly JT active). Use of this local basis is more convenient
for analysis of the electronic states and occupancies. The transformation
from global to local coordinates is performed for each relaxed geometry
by employing a direct polynomial-based transformation of orbitals
(detailed in Appendix B). Unless specifically noted, orbitals and
occupancies refer to the local basis.}

\section{Results and Discussion}

At $750$ K LaMnO\textsubscript{%
3%
} (LMO) undergoes a structural phase transition, transforming from
cubic to orthorhombic symmetry. Under ambient conditions the orthorhombic
perovskite has a paramagnetic spin structure. Below the N\'eel temperature
of\emph{ }$T_{\text{N}}\thickapprox140$ K, \citep{Kovaleva2004}
LMO is an insulator with A-type antiferromagnetic (A-AFM) spin ordering.
In the low\emph{ T} orthorhombic P\emph{nma }LMO shown in Figure \ref{fig:exptlmostruct},
experimental reports of lattice parameters are $a=5.736\ {\rm \AA}$,
$b=7.703\ {\rm \AA}$ and $c=5.540\ {\rm \AA}$ by neutron powder
diffraction\citep{SakaiN.;FjellvagH.;Lebech1997}. To support the
A-AFM ordering in LaMnO$_{3}$, the Mn \emph{d}\textsuperscript{%
4%
} electrons exchange anisotropically: ferromagnetic (FM) coupling exists
between Mn in $\{010\}$ planes while AFM coupling exists between
successive planes along $[010]$. 

The reported experimental band gaps in LMO cover a range of values,
depending on whether the gap is determined from measurements on conductivity
(0.24 eV),\citep{Mahendiran1996} optical absorption (1.1 eV),\citep{Arima1993}
photoemission (1.7 eV),\citep{Saitoh1995} optical conductivity (1.9
eV),\citep{Jung1997} or resonant Raman spectroscopy (2 eV).\citep{Kruger2004}
DFT is a single-particle theory, so even with the exact exchange-correlation
functional, it can only describe the fundamental (quasiparticle) band
gap and not the optical one. We therefore consider the most appropriate
reference value to be the $1.7$ eV photoemission gap measured by
Saitoh \emph{et al}.\citep{Saitoh1995}. We note that recent computational
work by Lee \emph{et al}.\citep{Lee2013} predicts a direct gap of
1.1 eV, in agreement with the optical absorption gap of 1.1 eV measured
by Arima \emph{et al.}\citep{Arima1993}. The value of the optical
gap is generally lower than the fundamental gap due to electron-hole
interactions (\emph{i.e}., excitonic effects). Such two-particle interactions
are not included in standard one-particle DFT, so we believe the most
reliable comparisons should be made between a benchmark indirect experimental
photoemission gap such as the 1.7 eV Saitoh gap\citep{Saitoh1995}
and the indirect DFT gap.

One of our main practical considerations here is to reproduce the
different facets of the above experimental description. To do this,
DFT calculations are performed screening through different levels
of Coulombic localisation.

\subsection{Description of LaMnO\textsubscript{%
3%
} using \textcolor{black}{DFT+}\textcolor{black}{\emph{U}}}

Previous work has applied the single term $U_{\text{eff}}$ approach
to calculations on bulk LaMnO\textsubscript{%
3%
}.\citep{Trimarchi2005,Sawada1997,Franchini,Chen2012} The failure
of this approach to \emph{simultaneously} describe the energy gap,
structure and magnetism drives us to systematically examine the $U_{\text{eff}}$
method. These initial results also provide context for the more sophisticated
\emph{$U|J$} method and analysis of its merits and behaviour below.

\subsubsection{\emph{Experimental LaMnO$_{3}$ structure }via \textcolor{black}{\emph{DFT+}}\textcolor{black}{U}\textcolor{black}{\emph{\textsubscript{%
\textcolor{black}{eff}%
}}}}

\begin{table}
\caption{%
\label{table:bulkdudarev}LDA+$U_{\text{eff}}$ and GGA+$U_{\text{eff}}$
results for the energy gap $E^{\text{Gap}}$ (in eV), for the A-AFM
and FM phases, and the total energy difference $\lyxmathsym{\textgreek{D}}E=E^{\text{A-AFM}}-E^{\text{FM}}$
(in meV) per formula unit of LaMnO$_{3}$. The crystal structure is
held fixed at the experimental geometry.%
}

\begin{tabular}{cccccccc}
\hline 
\noalign{\vskip\doublerulesep}
\multirow{2}{*}{$U_{\text{eff}}$\emph{ }(eV)} & \multicolumn{4}{c}{LDA (PZ81)} & \multicolumn{3}{c}{GGA (PBEsol)}\tabularnewline[\doublerulesep]
\cline{2-4} \cline{6-8} 
\noalign{\vskip\doublerulesep}
 & $E_{\text{A-AFM}}^{\text{Gap}}$ & $\Delta_{\text{FM}}^{\text{Gap}}$ & $\lyxmathsym{\textgreek{D}}E$  & %
 & $E_{A-\text{AFM}}^{\text{Gap}}$ & $\Delta_{\text{FM}}^{\text{Gap}}$ & $\lyxmathsym{\textgreek{D}}E$ \tabularnewline[\doublerulesep]
\hline 
\hline 
\noalign{\vskip\doublerulesep}
0 & 0.0 & 0.0 & -22 & %
 & 0.2 & 0.0 & -13\tabularnewline[\doublerulesep]
\noalign{\vskip\doublerulesep}
2 & 0.5 & 0.0 & -5 & %
 & 0.6 & 0.0 & 1\tabularnewline[\doublerulesep]
\noalign{\vskip\doublerulesep}
4 & 1.0 & 0.0 & 4 & %
 & 1.0 & 0.0 & 8\tabularnewline[\doublerulesep]
\noalign{\vskip\doublerulesep}
6 & 1.3 & 0.1 & 10 & %
 & 1.3 & 0.1 & 14\tabularnewline[\doublerulesep]
\noalign{\vskip\doublerulesep}
8 & 1.4 & 0.2 & 14 & %
 & 1.4 & 0.2 & 17\tabularnewline[\doublerulesep]
\hline 
\end{tabular}
\end{table}

Standard LDA (PZ81) and GGA (PBEsol) with $U_{\text{eff}}=0$ eV both
successfully stabilize the low temperature experimental A-AFM ordering
as shown in Table \ref{table:bulkdudarev}. However this is essentially
where the success ends. As noted previously, both GGA and LDA are
often unable to produce significant orbital splitting (beyond some
aspects due to spin exchange and structural distortion) and also exaggerate
electron delocalization due to inexact exchange (or equivalently lack
of self-interaction correction). This inevitably results in a qualitatively
incorrect electronic structure with a seriously underestimated band
gap: both GGA and LDA with\emph{ }$U_{\text{eff}}$ = 0 eV yield band
gaps that are far too small compared to experiment.

\textcolor{black}{Increasing }$U_{\text{eff}}$\textcolor{black}{{}
stabilizes the occupied $(f_{i\sigma}\apprge\frac{1}{2}$) eigenstates
and drives orbital occupations toward binary polarization: filled
states become more filled and empty states more empty.}\textcolor{green}{{}
}\textcolor{black}{For example, increasing }$U_{\text{eff}}$\textcolor{black}{{}
from 0 to 8 eV in GGA calculations results in the following change
in occupancies in the Mn }\textcolor{black}{\emph{d}}\textcolor{black}{{}
manifold:}

\textcolor{black}{
\begin{equation}
(f_{\sigma}|f_{\bar{\sigma}})=\left(\begin{array}{c|c}
0.65 & 0.22\\
0.73 & 0.26\\
0.93 & 0.11\\
0.93 & 0.10\\
0.93 & 0.09
\end{array}\right)\rightarrow\left(\begin{array}{c|c}
1.00 & 0.10\\
0.57 & 0.17\\
0.98 & 0.04\\
0.97 & 0.03\\
0.97 & 0.04
\end{array}\right),\label{eq: SA occupancy change}
\end{equation}
where the ordering of orbitals in the local basis is $\left(\begin{array}{c}
3z^{2}-r^{2}\\
x^{2}-y^{2}\\
xy\\
yz\\
xz
\end{array}\right)$. }

The Hubbard limit of very large\emph{ }$U_{\text{eff}}$ typically
favours FM coupling in\emph{ }LaMnO\textsubscript{%
3%
},\citep{Millis1997} and Table \ref{table:bulkdudarev} confirms this.
The primary reason is that increasing $U_{\text{eff}}$ kills the
superexchange mechanism, which scales as $\sim t^{2}/U_{\text{eff}}$
where $t$ is the effective Mn-Mn hopping, and this mechanism underlies
the stability of A-AFM ordering in LaMnO\textsubscript{%
3%
}. As a result, \emph{e\textsubscript{%
\emph{g}%
}} double-exchange is relatively strengthened and we find FM ordering.
Critically, Table \ref{table:bulkdudarev} shows that a large $U_{\text{eff}}$
value is required to open a satisfactory energy gap. Unfortunately
this situation results in a trade-off between correct gap or correct
magnetism.

\subsubsection{\emph{Relaxed LaMnO$_{3}$ structure }via\emph{ DFT+$U_{\text{eff}}$}}

\begin{table*}
\caption{%
Band gap $E^{\text{Gap}}$\emph{,} total energy difference $\lyxmathsym{\textgreek{D}}E=E^{\text{A-AFM}}-E^{\text{FM}}$
per unit cell, and percent errors, with respect to experiment, for
lattice parameters and unit cell volume of fully relaxed A-AFM bulk
LaMnO$_{3}$.\label{table:dudarevrelaxed}%
}

\begin{tabular}{cccccccccccccc}
\hline 
\noalign{\vskip\doublerulesep}
\multirow{2}{*}{$U_{\text{eff}}$ (eV)} & \multicolumn{7}{c}{LDA (PZ81)} & \multicolumn{6}{c}{GGA (PBEsol)}\tabularnewline[\doublerulesep]
\cline{2-7} \cline{9-14} 
\noalign{\vskip\doublerulesep}
 & $E^{\text{Gap}}$ (eV) & $\Delta^{\text{a}}$ (\%) & $\Delta^{\text{b}}$ (\%) & $\Delta^{\text{c}}$ (\%) & $\Delta^{\text{Vol.}}$ (\%) & \textcolor{black}{\emph{$\Delta E$ }}\textcolor{black}{(meV)} & %
 & \textcolor{black}{$E^{\text{Gap}}$}\textcolor{black}{\emph{ }}\textcolor{black}{(eV)} & $\Delta^{\text{a}}$ (\%) & $\Delta^{\text{b}}$ (\%) & $\Delta^{\text{c}}$ (\%) & $\Delta^{\text{Vol.}}$ (\%) & \textcolor{black}{\emph{$\Delta E$ }}\textcolor{black}{(meV)}\tabularnewline[\doublerulesep]
\hline 
\hline 
\noalign{\vskip\doublerulesep}
0 & 0.00 & -5.8 & -3.0 & -1.2 & -9.8 & \textcolor{black}{54 } & %
 & \textcolor{black}{0.00} & \textcolor{black}{-3.5} & \textcolor{black}{-0.6} & \textcolor{black}{-0.3} & \textcolor{black}{-4.4} & \textcolor{black}{34}\tabularnewline[\doublerulesep]
\noalign{\vskip\doublerulesep}
2 & 0.22 & -2.9 & -1.6 & -1.1 & -5.5 & \textcolor{black}{52 } & %
 & \textcolor{black}{0.48} & \textcolor{black}{-0.6} & \textcolor{black}{-0.8} & \textcolor{black}{-0.5} & \textcolor{black}{-1.9} & \textcolor{black}{47}\tabularnewline[\doublerulesep]
\noalign{\vskip\doublerulesep}
4 & 0.81 & -1.8 & -1.6 & -1.1 & -4.4 & \textcolor{black}{25} & %
 & \textcolor{black}{0.92} & \textcolor{black}{-0.0} & \textcolor{black}{-0.5} & \textcolor{black}{-0.3} & \textcolor{black}{-0.9} & \textcolor{black}{19}\tabularnewline[\doublerulesep]
\noalign{\vskip\doublerulesep}
6 & 1.13 & -1.5 & -1.0 & -1.0 & -3.5 & \textcolor{black}{13 } & %
 & \textcolor{black}{1.10} & \textcolor{black}{0.2} & \textcolor{black}{-0.1} & \textcolor{black}{-0.3} & \textcolor{black}{-0.1} & \textcolor{black}{19}\tabularnewline[\doublerulesep]
\noalign{\vskip\doublerulesep}
8 & 1.23 & -1.4 & -0.6 & -1.0 & -3.0 & \textcolor{black}{19} & %
 & \textcolor{black}{1.08} & \textcolor{black}{0.4} & \textcolor{black}{0.4} & \textcolor{black}{-0.2} & \textcolor{black}{0.6} & \textcolor{black}{27}\tabularnewline[\doublerulesep]
\hline 
\end{tabular}
\end{table*}

\begin{figure*}
\includegraphics[scale=0.33]{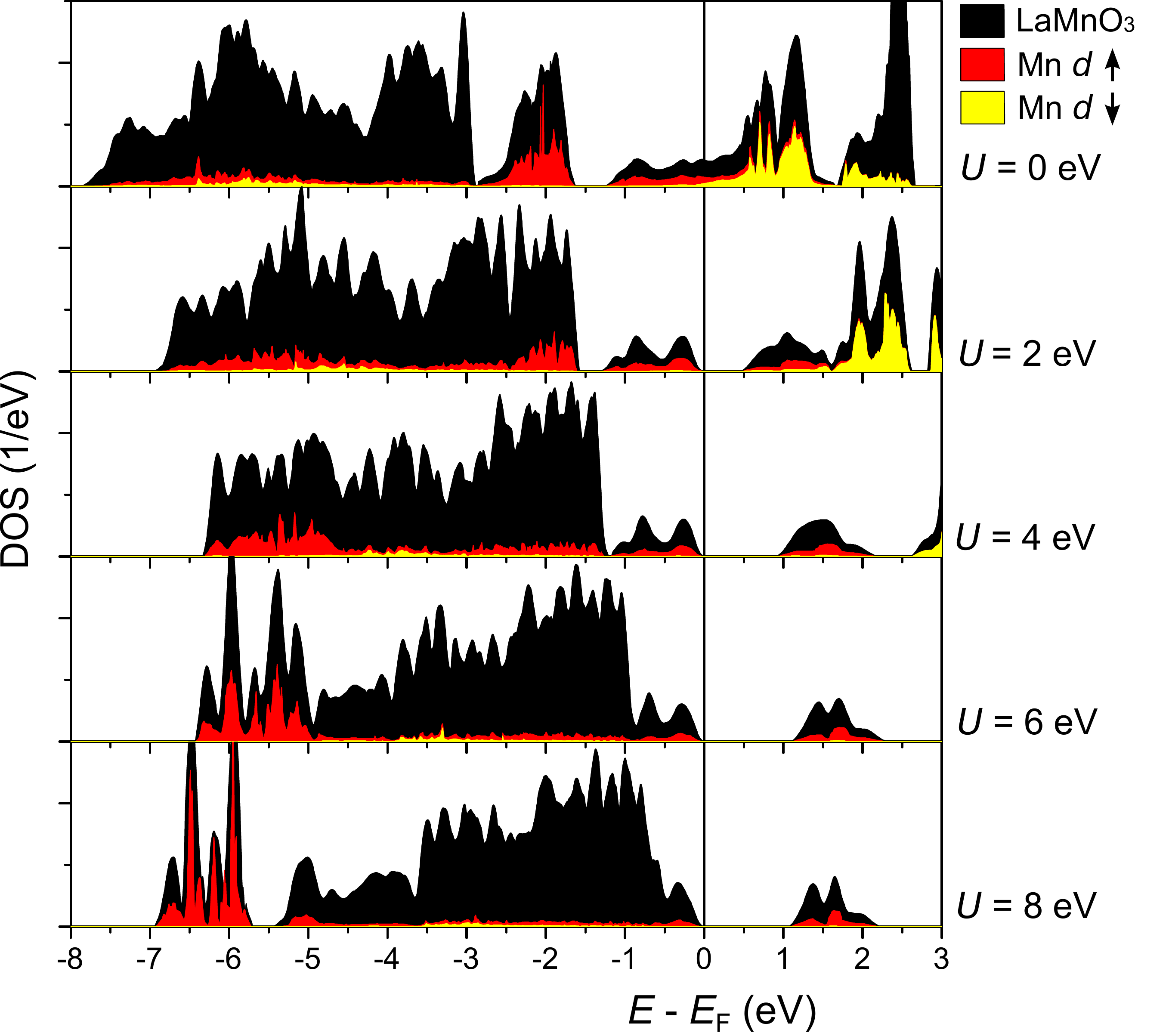} \includegraphics[scale=0.35]{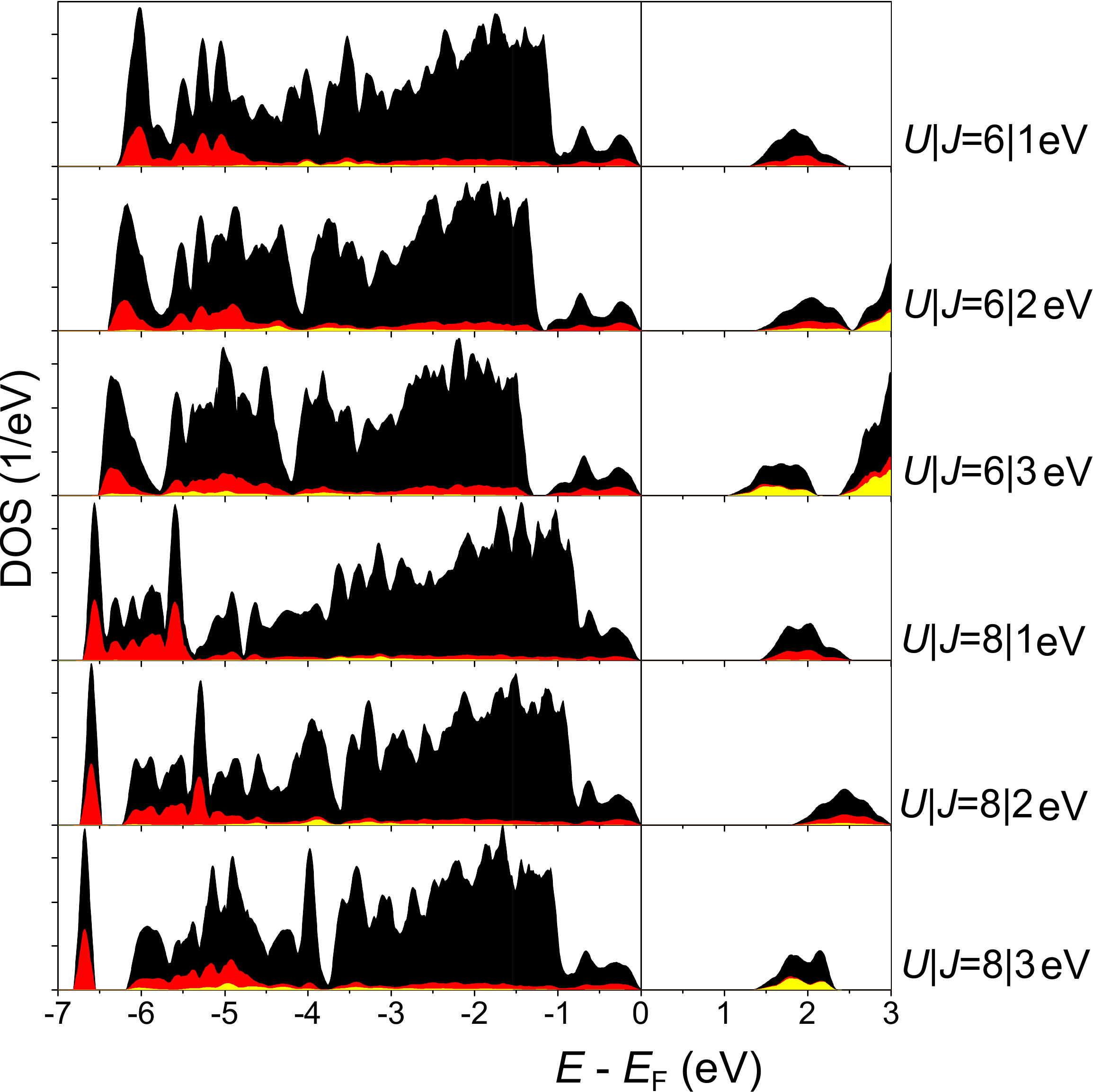}

\caption{%
\textcolor{black}{$U_{\text{eff}}$}\textcolor{black}{\emph{ }}\textcolor{black}{(left)}\textcolor{black}{\emph{
}}\textcolor{black}{and}\textcolor{black}{\emph{ $U|J$ }}\textcolor{black}{(right)
corrected density of states for fully relaxed A}\textcolor{black}{\emph{-}}\textcolor{black}{AFM
LaMnO$_{3}$, as a function of energy, $E-E_{\text{F}}$.} Black curves
show the total density of states while red and yellow curves show
Mn \textit{d} majority spin and minority spin densities of states.\label{fig:combinedDOS}\textcolor{green}{{} }%
}
\end{figure*}

When we permit the structure of LaMnO\textsubscript{%
3%
} to fully relax during the calculation, we find the results in Table
\ref{table:dudarevrelaxed}. We see that having a non-zero $U_{\text{eff}}$
improves the crystal geometry and the electronic structure description
for both GGA and LDA. Particular improvements are for the large erroneous
distortion in \emph{a }(insufficient orthorhombic character) and the
opening of the band gap. Figure \textcolor{black}{\ref{fig:combinedDOS}
}and Table \ref{table:dudarevrelaxed} show that the band gap increases
roughly linearly with $U_{\text{eff}}$ at first but then tails off
at higher $U_{\text{eff}}$. \textcolor{black}{The ineffectiveness
of $U_{\text{eff}}$ at high values is shown in Figure \ref{fig:combinedDOS},
and can be understood in terms of the partial $x^{2}-y^{2}$ occupation
shown in Equation (\ref{eq: SA occupancy change}). The partial $x^{2}-y^{2}$
occupation damps}\textcolor{black}{\emph{ }}\textcolor{black}{the
impact of }$U_{\text{eff}}$\textcolor{black}{{} on the energy eigenvalue
splittings since $\Delta\epsilon_{x^{2}-y^{2}\sigma}=U_{\text{eff}}\left(\frac{1}{2}-f_{x^{2}-y^{2}\sigma}\right)\approx0$
for $f_{x^{2}-y^{2}\sigma}\approx\frac{1}{2}$. (The reason partial
$e_{\text{g}}$ occupation occurs is that the }\textcolor{black}{\emph{d}}\textcolor{black}{{}
manifold is not isolated but connected to the rest of the system }\textcolor{black}{\emph{via}}\textcolor{black}{{}
hybridization to the O $2$}\textit{\textcolor{black}{p}}\textcolor{black}{{}
orbitals, or in other words due to the partial covalency of the Mn-O
bond.) }

Although adding\emph{ }$U_{\text{eff}}$ to GGA and LDA produces similar
band gaps as per Table \ref{table:dudarevrelaxed}, the GGA+$U_{\text{eff}}$
geometry is preferable. Overall a high value of $U_{\text{eff}}$
$\sim$ $6$ eV, correcting the GGA formalism, provides on balance
the best gap/structure combination.\textcolor{black}{{} Again, an evident
failure of $U_{\text{eff}}$ is its inability to predict the correct
magnetic ordering at the $U_{\text{eff}}$}\textcolor{black}{\emph{
}}\textcolor{black}{level required to correct the structure and the
band gap}\textcolor{black}{\emph{.}} 

\subsubsection{\emph{Experimental LaMnO$_{3}$ structure }via \emph{DFT+$U|J$} }

\begin{table}
\caption{%
Results from GGA (PBEsol) + $U|J$ for the experimental geometry of
bulk LaMnO\textsubscript{3}. Band gaps $E^{\text{Gap}}$ are in eV
for each magnetic state, and $\lyxmathsym{\textgreek{D}}E=E^{\text{A-AFM}}-E^{\text{FM}}$
is the total energy difference per unit cell between the two magnetic
phases. \label{table:UJexptbulk}%
}

\begin{tabular}{cccc}
\hline 
\noalign{\vskip\doublerulesep}
\textcolor{black}{\emph{$U|J$}}\textcolor{black}{{} (eV)} & \textcolor{black}{$E_{\text{A-AFM}}^{\text{Gap}}$ (eV)} & \textcolor{black}{$E_{\text{FM}}^{\text{Gap}}$ (eV)} & \textcolor{black}{\emph{$\Delta E$ }}\textcolor{black}{(meV)}\tabularnewline[\doublerulesep]
\hline 
\hline 
\noalign{\vskip\doublerulesep}
\textcolor{black}{$6|0$} & \textcolor{black}{1.3} & \textcolor{black}{0.1} & \textcolor{black}{14}\tabularnewline[\doublerulesep]
\noalign{\vskip\doublerulesep}
\textcolor{black}{$6|1$} & \textcolor{black}{1.3} & \textcolor{black}{0.2} & \textcolor{black}{6}\tabularnewline[\doublerulesep]
\noalign{\vskip\doublerulesep}
\textcolor{black}{$6|2$} & \textcolor{black}{1.2} & \textcolor{black}{0.4} & \textcolor{black}{-10}\tabularnewline[\doublerulesep]
\noalign{\vskip\doublerulesep}
\textcolor{black}{$6|3$} & \textcolor{black}{0.6} & \textcolor{black}{0.2} & \textcolor{black}{-39}\tabularnewline[\doublerulesep]
\noalign{\vskip\doublerulesep}
\textcolor{black}{$8|0$} & \textcolor{black}{1.4} & \textcolor{black}{0.2} & \textcolor{black}{17}\tabularnewline[\doublerulesep]
\noalign{\vskip\doublerulesep}
\textcolor{black}{$8|1$} & \textcolor{black}{1.5} & \textcolor{black}{0.4} & \textcolor{black}{11}\tabularnewline[\doublerulesep]
\noalign{\vskip\doublerulesep}
\textcolor{black}{$8|2$} & \textcolor{black}{1.6 } & \textcolor{black}{0.8} & \textcolor{black}{-2}\tabularnewline[\doublerulesep]
\noalign{\vskip\doublerulesep}
\textcolor{black}{$8|3$} & \textcolor{black}{1.0} & \textcolor{black}{0.5} & \textcolor{black}{-32}\tabularnewline[\doublerulesep]
\hline 
\end{tabular}
\end{table}

Following the failure of the $U_{\text{eff}}$ scheme in both single-point
and relaxed geometry calculations, we turn to the DFT+\emph{$U|J$}
methodology. The $U_{\text{eff}}$ results conveniently suggest a
reasonable starting point: since $U_{\text{eff}}=U-J$\emph{,} an
\emph{$U|J$} correction with magnitude of approximately $U-J\approx6$
eV is appropriate. Results in Table \ref{table:UJexptbulk} are for
bulk LaMnO\textsubscript{%
3 %
}at the experimental structure, and sample \emph{J} from 0 to 3 eV
in conjunction with $U=6$ eV and $8$ eV. Increasing \emph{J} for
a fixed value of \emph{U} stabilizes A-AFM ordering and enhances orbital
splitting which further opens the band gap. Orbital splittings due
to the $U_{\text{eff}}$ are generally ``isotropic'' in that they
are based solely on the occupation. The marked improvement by the
\emph{$U|J$} method emphasises the importance of explicit spatial
exchange anisotropy in the LaMnO$_{3}$ Mn \emph{d} manifold. The
results in Table \ref{table:UJexptbulk} are encouraging, but since
distortion of the lattice is critical in LaMnO\textsubscript{%
3%
},\citep{Solovyev1996} the trends observed must be verified by fully
relaxing ionic positions and lattice parameters, which we report on
next.

\subsubsection{\emph{Relaxed LaMnO$_{3}$ structure} via\emph{ DFT+$U|J$} }

\begin{table*}
\caption{%
Fully relaxed LaMnO$_{3}$ results based on GGA (PBEsol) + \emph{$U|J$}.
Band gaps $E^{\text{Gap}}$, lattice parameter errors, and total energy
differences between the A-AFM and FM magnetic phases $\lyxmathsym{\textgreek{D}}E=E^{\text{A-AFM}}-E^{\text{FM}}$
per formula unit are listed.\label{table:UJrelaxed}%
}

\begin{tabular}{cccccccc}
\hline 
\noalign{\vskip\doublerulesep}
\multirow{2}{*}{\emph{$U|J$} (eV)} & \multicolumn{6}{c}{A-AFM} & \multirow{2}{*}{$\lyxmathsym{\textgreek{D}}E$\emph{ }(meV)}\tabularnewline[\doublerulesep]
\cline{2-6} 
\noalign{\vskip\doublerulesep}
 & $E^{\text{Gap}}$ (eV) & $\Delta^{\text{a}}$ (\%) & $\Delta^{\text{b}}$ (\%) & $\Delta^{\text{c}}$ (\%) & $\Delta^{\text{Vol.}}$ (\%) & %
 & \tabularnewline[\doublerulesep]
\hline 
\hline 
\noalign{\vskip\doublerulesep}
\textcolor{black}{$6|0$} & \textcolor{black}{1.1} & \textcolor{black}{0.4} & \textcolor{black}{-0.1} & \textcolor{black}{-0.4} & \textcolor{black}{0.0} & %
 & \textcolor{black}{19}\tabularnewline[\doublerulesep]
\noalign{\vskip\doublerulesep}
$6|0.5$ & 1.2 & 0.6 & -0.4 & -0.4 & -0.2 & %
 & 14\tabularnewline[\doublerulesep]
\noalign{\vskip\doublerulesep}
\textcolor{black}{$6|1$} & \textcolor{black}{1.3} & \textcolor{black}{0.9} & \textcolor{black}{-0.7} & \textcolor{black}{-0.5} & \textcolor{black}{-0.3} & %
 & \textcolor{black}{7}\tabularnewline[\doublerulesep]
\noalign{\vskip\doublerulesep}
$6|1.5$ & 1.4 & 1.2 & -0.9 & -0.6 & -0.3 & %
 & -1\tabularnewline[\doublerulesep]
\noalign{\vskip\doublerulesep}
\textcolor{black}{$6|2$} & \textcolor{black}{1.4} & \textcolor{black}{0.8} & \textcolor{black}{-1.2} & \textcolor{black}{0.1} & \textcolor{black}{-0.4} & %
 & \textcolor{black}{-10}\tabularnewline[\doublerulesep]
\noalign{\vskip\doublerulesep}
$6|2.5$ & 1.3 & 1.8 & -1.3 & -0.8 & -0.2 & %
 & -21\tabularnewline[\doublerulesep]
\cline{2-8} 
\noalign{\vskip\doublerulesep}
$7|0$ & 1.2 & 0.5 & 0.1 & -0.3 & 0.3 & %
 & 23\tabularnewline[\doublerulesep]
\noalign{\vskip\doublerulesep}
$7|0.5$ & 1.3 & 0.7 & -0.2 & -0.4 & 0.2 & %
 & 17\tabularnewline[\doublerulesep]
\noalign{\vskip\doublerulesep}
$7|1$ & 1.4 & 0.9 & -0.5 & -0.5 & 0.0 & %
 & 11\tabularnewline[\doublerulesep]
\noalign{\vskip\doublerulesep}
$7|1.5$ & 1.5 & 1.2 & -0.7 & -0.5 & -0.1 & %
 & 3\tabularnewline[\doublerulesep]
\noalign{\vskip\doublerulesep}
$7|2$ & 1.6 & 1.6 & -1.0 & -0.7 & 0.0 & %
 & -6\tabularnewline[\doublerulesep]
\noalign{\vskip\doublerulesep}
$7|2.5$ & 1.5 & 1.9 & -1.2 & -0.7 & -0.1 & %
 & -17\tabularnewline[\doublerulesep]
\cline{2-8} 
\noalign{\vskip\doublerulesep}
\textcolor{black}{$8|0$} & \textcolor{black}{1.1} & \textcolor{black}{0.5} & \textcolor{black}{0.4} & \textcolor{black}{-0.3} & \textcolor{black}{0.6} & %
 & \textcolor{black}{27}\tabularnewline[\doublerulesep]
\noalign{\vskip\doublerulesep}
$8|0.5$ & 1.2 & 0.6 & 0.1 & -0.3 & 0.4 & %
 & 21\tabularnewline[\doublerulesep]
\noalign{\vskip\doublerulesep}
\textcolor{black}{$8|1$} & \textcolor{black}{1.4} & \textcolor{black}{1.0} & \textcolor{black}{-0.2} & \textcolor{black}{-0.5} & \textcolor{black}{0.3} & %
 & \textcolor{black}{15}\tabularnewline[\doublerulesep]
\noalign{\vskip\doublerulesep}
$8|1.5$ & 1.6 & 1.2 & -0.5 & -0.5 & 0.1 & %
 & 7\tabularnewline[\doublerulesep]
\noalign{\vskip\doublerulesep}
\textcolor{black}{$8|2$} & \textcolor{black}{1.8} & \textcolor{black}{1.5} & \textcolor{black}{-0.8} & \textcolor{black}{-0.6} & \textcolor{black}{0.1} & %
 & \textcolor{black}{-2}\tabularnewline[\doublerulesep]
\noalign{\vskip\doublerulesep}
$8|2.5$ & 1.7 & 1.7 & -1.0 & --0.6 & 0.1 & %
 & -14\tabularnewline[\doublerulesep]
\hline 
\end{tabular}
\end{table*}

Table \ref{table:UJrelaxed} displays key data for fully relaxed bulk
LaMnO\textsubscript{%
3%
} using the DFT+\emph{$U|J$} framework. Relaxed results largely echo
the experimental structure results above for A-AFM LaMnO$_{3}$, with
the \emph{$U|J$} combination of \emph{$U=8$} eV and \emph{$J=2$}
eV providing a good material description. In particular the $U|J=8|2$
eV combination provides agreement in terms of electronic, magnetic
and structural observables from experiment\citep{SakaiN.;FjellvagH.;Lebech1997,Elemans1971,Saitoh1995,Jung1997}
and also more computationally expensive many body GW approximation\citep{Nohara2008}
results. Volume errors $<1$ \% improve on previous work,\citep{Sawada1997,Elemans1971,SakaiN.;FjellvagH.;Lebech1997}
and the error in energy gap is small at approximately $\sim5$ \%
($<0.1$ eV error)\citep{Saitoh1995} In addition the A-AFM ordering
is stabilized against FM ordering which was previously seen as a missing
ingredient\citep{Sawada1997,Hashimoto2010}. The improvements in LMO
description depend intimately on the intra-orbital exchange description
- this is explored further by quantifying the action of the Hund's
coupling interaction on the LMO Mn \emph{d} states.

\subsection{Explicit exchange anisotropy in Mn\textsuperscript{%
3+%
}}

Strong on-site Coulomb repulsion is the central theme in paradigms
of ``Mottness'' and electron localisation. However, the importance
of Hund's coupling (intra-orbital exchange) in materials with partial
\emph{d }and \emph{f-}shell\emph{ }occupations has been highlighted.\citep{Georges2013}
In this section, we attempt to understand the nature of Hund's coupling
in LMO, by examining the effects of the on-site exchange terms as
defined in Appendix A. We explore why the \emph{$U|J$} methodology
can describe LaMnO$_{3}$ adequately, reproducing band gap and correct
magnetic ground state simultaneously. We employ a simple model where
we focus only on the occupancies of the Mn\textsuperscript{%
3+%
} \emph{d}\textsuperscript{%
4%
} manifold in order to isolate the effect of the exchange \emph{J}
parameter (and related physics) on the Mn \emph{d} states as per Equation
(\ref{eq:energyshiftRIUJformula}). Majority spin \emph{t}\textsubscript{%
2g%
}\textsuperscript{%
3%
} states are generally fully occupied due to the strong exchange splitting
between spin channels, and as is well known, increasing \emph{U} increases
occupancy polarization. However the nature of anisotropic interactions
in the Mn \emph{d} shell due to \emph{J} is less well understood,
particularly with respect to the polarization of the \emph{e}\textsubscript{%
g%
}\textsuperscript{%
1%
} occupation into $3z^{2}-r^{2}$ or $x^{2}-y^{2}$ (or some mix of
the two).

We begin with three model \emph{e}\textsubscript{%
g%
}\textsuperscript{%
1%
} occupations, $\pi^{e_{\text{g}}}=0,\pm1$, in an attempt to pinpoint
what \emph{J} really does and understand the nature of Hund's coupling
in different limits. As a reminder, $\pi^{e_{\text{g}}}$ is the \emph{e}\textsubscript{%
g%
} occupancy polarization as defined in Equation (\ref{eq:orbital_polarisation_definition}).
After examining these model systems, we will consider the effect of
\emph{J} in the actual calculations where we use the calculated \emph{ab
initio} occupations together with our analytical rewriting of the
\emph{$U|J$} energy function and eigenvalue corrections. As explained
above, the\emph{ e}\textsubscript{%
g%
} and \emph{t}\textsubscript{%
2g%
} group terms discussed correspond to the local octahedral basis \textcolor{black}{(}\textcolor{black}{\emph{i.e.}}\textcolor{black}{,
post rotation as per Appendix }B).

\subsubsection{Anisotropic exchange for model orbital occupations}

To illustrate the anisotropic effects of the\textit{ J} terms in the
$U|J$ schema, we begin with a set of model occupancies where we fix
the formal occupation of Mn\textsuperscript{%
3+%
} (\textit{d}\textsuperscript{%
4%
}) but vary the orbital polarization. 

A $\pi^{e_{\text{g}}}=+1$ model polarization corresponds t\textcolor{black}{o
a single hole on the majority spin }$3z^{2}-r^{2}$ site (\emph{i.e.},
$f_{x^{2}-y^{2}\sigma}=f_{t_{\text{2g}}\sigma}=1$, $f_{3z^{2}-r^{2}\sigma}=0$
and $f_{\bar{\sigma}}=0$). Based on Equation (\ref{eq:energyshiftRIUJformula}),
the added effect of the exchange \textit{J }terms is to create additional
energy splittings (beyond simple occupancy polarization proportional
to $U_{\text{eff}}$) given by

\textcolor{black}{
\begin{equation}
\pi^{e_{\text{g}}}=+1\,\,:\,\,\,\,\,\,\,(\bigtriangleup\epsilon_{\sigma}|\triangle\epsilon_{\bar{\sigma}})=J\cdot\left(\begin{array}{c|c}
0.00 & -1.14\\
0.52 & 0.63\\
0.52 & 0.63\\
-0.52 & -0.06\\
-0.52 & -0.06
\end{array}\right).\label{eq: mod_pol_1}
\end{equation}
}The opposite polarity, $\pi^{e_{\text{g}}}=-1$, corresponds \textcolor{black}{to
a single hole on th}e majority spin $x^{2}-y^{2}$ site (that is,
$f_{3z^{2}-r^{2}\sigma}=f_{t_{\text{2g}}\sigma}=1$, $f_{x^{2}-y^{2}\sigma}=0$
and $f_{\bar{\sigma}}=0$). This results in the following exchange
energy splittings

\textcolor{black}{
\begin{equation}
\pi^{e_{\text{g}}}=-1\,\,:\,\,\,\,\,\,\,(\bigtriangleup\epsilon_{\sigma}|\triangle\epsilon_{\bar{\sigma}})=J\cdot\left(\begin{array}{c|c}
0.52 & 0.63\\
0.00 & -1.14\\
-0.86 & -0.29\\
0.17 & 0.40\\
0.17 & 0.40
\end{array}\right).\label{eq: mod_pol_-1}
\end{equation}
}Removing the polarization, $\pi^{e_{\text{g}}}=0$, the single hole
is equally spread over the two majority spin $e_{\text{g}}$ sites
($f_{e_{\text{g}}\sigma}=0.5$ and $f_{t_{\text{2g}}\sigma}=1$ and
$f_{\bar{\sigma}}=0$). This leads to the splittings

\begin{figure*}
\centering{}\includegraphics[scale=0.7]{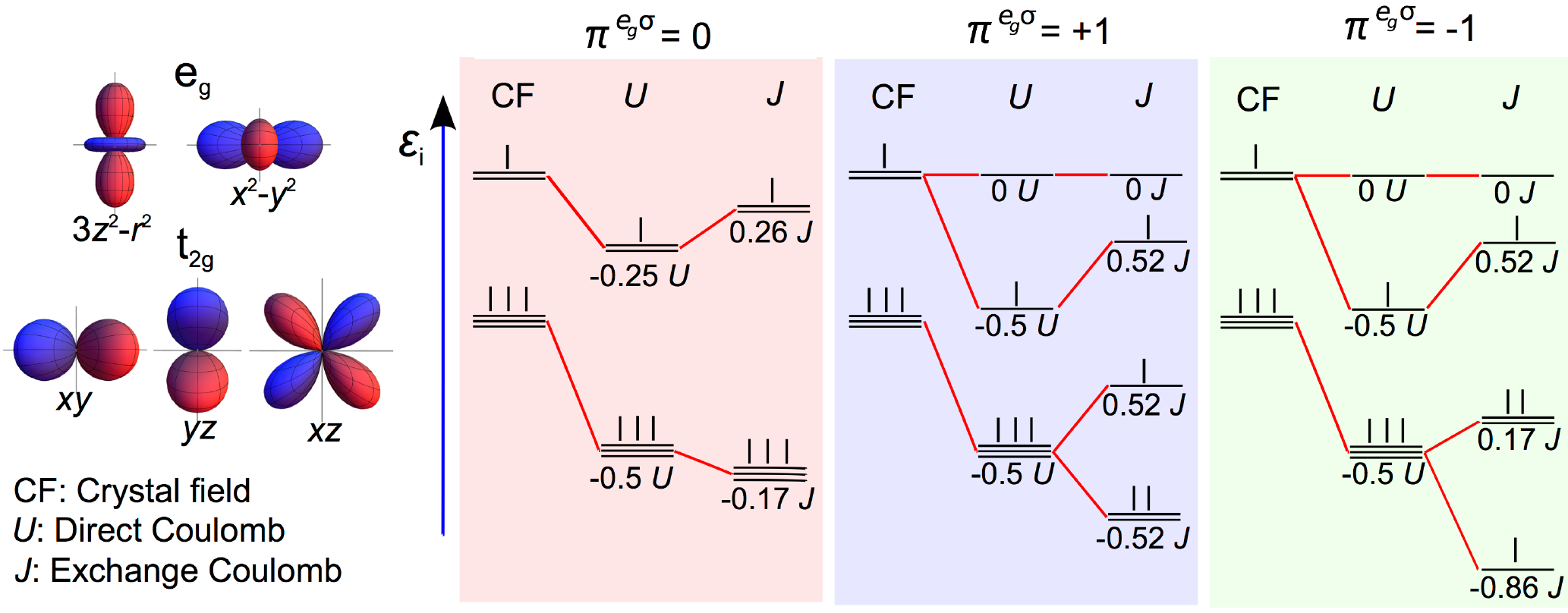}\caption{%
The occupation of states is represented in the model Mn\textsuperscript{3+}\emph{
d}\textsuperscript{4} manifold (majority spin only). Orbital degeneracy
is broken by octahedral crystal field (CF), Coulomb repulsion \emph{U}\textsubscript{eff}
(\textit{U }in the figure) and exchange \emph{J} following Equation
(\ref{eq:energyshiftRIUJformula}). Each vertical bar represents one
unit of electron occupation. $\pi^{e_{g}\sigma}$ is defined in Equation
(\ref{eq:orbital_polarisation_definition}) for which three limits
are examined: $\pi^{e_{g}\sigma}=0$ ($f_{x^{2}-y^{2}\sigma}=f_{3z^{2}-r^{2}\sigma}=0.5$),
$\pi^{e_{g}\sigma}=+1$ ($f_{x^{2}-y^{2}\sigma}=f_{3z^{2}-r^{2}\sigma}+1=1$),
and $\pi^{e_{g}\sigma}=-1$ ($f_{x^{2}-y^{2}\sigma}+1=f_{3z^{2}-r^{2}\sigma}=1$).
\label{fig:Schematic-of-modelJ}%
}
\end{figure*}

\textcolor{black}{
\begin{equation}
\pi^{e_{\text{g}}}=0\,\,:\,\,\,\,\,\,\,(\bigtriangleup\epsilon_{\sigma}|\triangle\epsilon_{\bar{\sigma}})=J\cdot\left(\begin{array}{c|c}
0.26 & -0.26\\
0.26 & -0.26\\
-0.17 & 0.17\\
-0.17 & 0.17\\
-0.17 & 0.17
\end{array}\right).\label{eq: mod_pol_0}
\end{equation}
}To visualize these results, we display a schematic showing these
splittings in these three cases of $\pi^{e_{\text{g}}}=0,\pm1$ in
Figure \ref{fig:Schematic-of-modelJ} where the corrections due to
both \textit{U} and \textit{J} are shown. 

These model results together with the Figure \ref{fig:Schematic-of-modelJ}
clearly point out that the effect of the \textit{J} terms is explicitly
anisotropic and its anisotropy and precise value depends on the orbital
polarization (which may have been present due the action of the $U_{\text{eff}}$
term). The anisotropy exists across both the magnetic quantum number
and spin channels ($\sigma$ and $\bar{\sigma}$). We now discuss
these three cases in more detail.

When we have full $e_{\text{g}}$ orbital polarization,\textit{ i.e.
}$\pi^{e_{\text{g}}}=\text{\textpm}1$, each polarity produces a unique
splitting pattern where the magnitude of anisotropy depends on the
sign of $\pi^{e_{\text{g}}}$. This is despite the fact that $3z^{2}-r^{2}$
and $x^{2}-y^{2}$ states both have $e_{\text{g}}$ symmetry: as we
can see that occupying each one (separately) splits the $t_{\text{2g}}$
quite differently. This difference is due to fact that the $x^{2}-y^{2}$
state is symmetry related to the $t_{\text{2}\text{g}}$ states (it
is the $xy$ state rotated by $\pi/4$ about the \textit{$z$} axis).
For example, when $x^{2}-y^{2}$ is fully occupied, the splittings
for $x^{2}-y^{2}$ and $xy$ are identical but differ from the other
orbitals, but the same is not true when $3z^{2}-r^{2}$ is filled
instead. Interestingly $\sum_{i\sigma}\Delta\epsilon_{i\sigma}f_{i\sigma}=0$
when $\pi^{e_{\text{g}}}=\text{\textpm}1$: this indicates that neither
polarization is energetically preferred by intra-orbital \emph{J}
terms. 

With zero $e_{\text{g}}$ orbital polarization,\textit{ i.e.} $\pi^{e_{\text{g}}}=0$,
we find that this degeneracy inhibits anisotropy from the \textit{J}
terms: the splitting \textit{within} each $t_{\text{2g}}$ and $e_{\text{g}}$
manifold is isotropic for both spin channels. The action of the \textit{J}
terms in this situation is to shift the energies of this manifold
\textit{en masse}. Here, $\sum_{i\sigma}\Delta\epsilon_{i\sigma}f_{i\sigma}=-0.26J$
when $\pi^{e_{\text{g}}}=0$ compared to zero for $\pi^{e_{\text{g}}}=\text{\textpm}1$\noun{.
}Hence, the anisotropic exchange terms in isolation actually favour
degenerate occupancy. This result appears to be counter-intuitive
given the importance of \textit{J} to anisotropy. The resolution is
that we have a much larger and dominant direct Coulomb term \textit{U}
that produces orbital polarization in the first place; the weaker
\textit{J }terms then further enlarge the polarization and make the
system more anisotropic. Table \ref{tab:Occupation-polarization-table}
shows this behaviour numerically. 

In brief, we see that \emph{J} acting alone\emph{ }energetically\emph{
favours degeneracy}. However, with a strong \textit{U} term already
creating orbital polarization, the \textit{J }terms provide the enlarged
anisotropic splitting that one finds in the final results of the calculation.

\subsubsection{Anisotropic exchange for ab initio orbital occupations}

\begin{figure*}
\centering{}\includegraphics[scale=0.2]{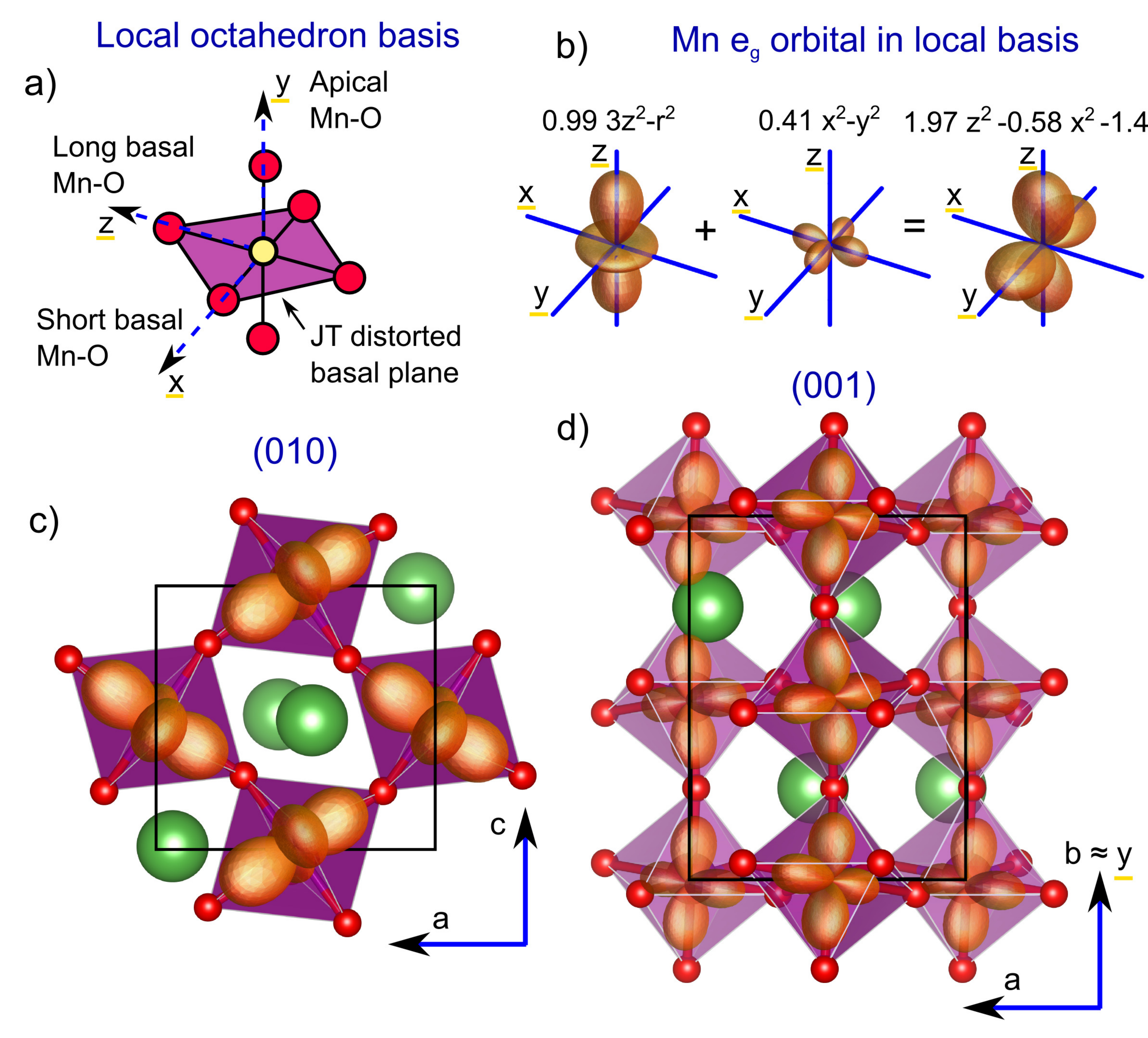}\caption{%
Orbitals in the LaMnO$_{3}$ unit cell from a \emph{$U|J=8|2$} eV
calculation. a) MnO\textsubscript{6} octahedron with Jahn-Teller
distorted plane and local basis vectors labelled. b) Visualization
of the occupation of the $3z^{2}-r^{2}$ and $x^{2}-y^{2}$ states
in the rotated basis of the density matrix as well as their superposition
for the total local $e_{\text{g}}$ occupancy (plotting the occupation
times the orbital expressed in spherical harmonics). c) The ordering
of the occupied $e_{\text{g}}$ shell ($1.97z^{2}-0.58x^{2}-1.4y^{2}$)
in the $(010)$ Jahn-Teller distorted FM coupled plane. d) The ordering
of the occupied $e_{\text{g}}$ shell in the $(001)$ plane with AFM
coupling along $b$. \label{fig:Orbital-ordering-across-LMO} Note
$x,y,z$ is the local octahedron basis, and $a,b,c$ lattice vectors
correspond to $x',y',z'$ global (pre-rotation) calculation basis. %
}
\end{figure*}

For the \emph{ab initio} orbital occupations we use the DFT+$U|J=8|2$\emph{
}eV calculation results, which yield an occupation-polarized $e_{\text{g}}$
manifold as shown in Table \ref{tab:Occupation-polarization-table}.
The $e_{\text{g}}$ polarity is found to be orbitally ordered across
the LaMnO$_{3}$ unit cell as shown in Figure \ref{fig:Orbital-ordering-across-LMO}.
We now examine this situation in more detail.

\textcolor{black}{The Mn }\textcolor{black}{\emph{d}}\textcolor{black}{{}
occupancies from the $U|J=8|2$ eV method with relaxed geometry are}

\begin{equation}
(f_{\sigma}|f_{\bar{\sigma}})=\left(\begin{array}{c|c}
0.99 & 0.10\\
0.41 & 0.33\\
0.98 & 0.06\\
0.96 & 0.04\\
0.97 & 0.05
\end{array}\right).\label{eq: RI occupation}
\end{equation}
These occupancies correspond to $\pi^{e_{\text{g}}\sigma}=-0.41$
(from $f_{3z^{2}-r^{2}\sigma}=0.99$ and $f_{x^{2}-y^{2}\sigma}=0.41$).\textcolor{black}{{}
}The Mn \emph{d} shell obviously has more electrons than the model
system above which was based on formal occupancies for Mn\textsuperscript{%
3+%
}. Again, we note that this increase is due to hybridization of the
Mn\textit{ d }orbitals with the neighboring O \textit{$2$p }orbitals
which admixes some Mn \textit{d} into the low-energy occupied valence
states and increases the electron count. Put differently, itineracy
due to the kinetic energy minimization competes with Hubbard-esque
Coulomb repulsion and we reach a balance. Numerically, for \textcolor{black}{$U|J=8|2$
eV,} the oxidation state based on the above Mn \textit{d} occupations
can be calculated to be $2.12+$ (an alternative or complementary
Bader charge picture yields an oxidation state of $1.68+$, still
less than the formal $3+$). 

The \textcolor{black}{$U|J=8|2$ eV} occupancies of Equation (\ref{eq: RI occupation})
result in energy splitting beyond splitting from $U_{\text{eff}}$
alone:

\textcolor{black}{
\begin{equation}
(\bigtriangleup\epsilon_{\sigma}|\triangle\epsilon_{\bar{\sigma}})=J\cdot\left(\begin{array}{c|c}
0.15 & 0.20\\
0.30 & -0.65\\
-0.41 & -0.06\\
-0.03 & 0.18\\
-0.01 & 0.20
\end{array}\right).\label{eq: uj82_energy_splitting}
\end{equation}
}The $e_{\text{g}}$ occupancy polarization of the \textcolor{black}{$U|J=8|2$
eV} calculation is considerably weaker than the previous model cases.
Nevertheless, it is large enough to drive significant anisotropic
exchange splittings in Equation (\ref{eq: uj82_energy_splitting}).
For example, the splittings are anisotropic within the majority spin
$t_{\text{2g}}$ manifold: the $xy$ state is pushed down by approximately
$0.4J$ compared to the other two $t_{\text{2g}}$ states. Within
the $e_{\text{g}}$ manifold, the fully occupied $3z^{2}-r^{2}$ state
is pushed up by $0.15J$ while the partially occupied $x^{2}-y^{2}$
state is pushed up considerably more by $0.30J$. 

The direct Coulomb interaction $U$ obviously increases $\pi^{e_{\text{g}}}$,
as expected from the standard instability condition for orbital polarization\citep{Sawada1997,Lu2009},
\[
U_{\text{eff}}\times D^{\sigma}(E_{\text{F}})\gg1\,,
\]
where $D^{\sigma}(E_{\text{F}})$ is the density of states in the
$\sigma$ spin channel at the Fermi level. The origin of the monotonic
relation between $\pi^{e_{\text{g}}}$ and \emph{J,} shown in Table
\ref{tab:Occupation-polarization-table}, is less obvious as $J$
is na\"ively expected to drive the electronic structure away from
orbital polarization as $U_{\text{eff}}=U-J$. However $\pi^{e_{\text{g}}}$
does increase with $J$, for the above-noted reason that $J$ alone
may favour orbital degeneracy but $J$ is strongly anisotropic when
in conjunction with a large $U$ value, resulting in the unequal upward
``push'' of the two $e_{\text{g}}$ orbitals with increasing $J$.
That $J$ and $\pi^{e_{\text{g}}}$ are so strongly coupled in this
material is interesting, as band gap, Jahn-Teller distortions, and
inter-site magnetic couplings all depend on $\pi^{e_{\text{g}}}$.

\begin{table}
\caption{%
Orbital occupation polarization, $\pi^{e_{\text{g}}\sigma}$, for
fully relaxed LaMnO\textsubscript{3} within the $U_{\text{eff}}$
and $U|J$ approaches. Majority spin are $\sigma$ and minority spin
are $\bar{\sigma}$. \label{tab:Occupation-polarization-table}%
}

\centering{}%
\begin{tabular}{ccc}
\hline 
\noalign{\vskip\doublerulesep}
\multirow{2}{*}{Correction (eV)} & Like-spin  & Opposite-spin \tabularnewline[\doublerulesep]
 & polarization, $\pi^{e_{g}\sigma}$  & polarization, $\pi^{e_{g}\bar{\sigma}}$\tabularnewline[\doublerulesep]
\hline 
\hline 
\noalign{\vskip\doublerulesep}
$U_{\text{eff}}=0$  & \textcolor{black}{$0.06$} & \textcolor{black}{$0.08$}\tabularnewline[\doublerulesep]
\noalign{\vskip\doublerulesep}
$U_{\text{eff}}=8$  & \textcolor{black}{$-0.27$} & \textcolor{black}{$0.25$}\tabularnewline[\doublerulesep]
\hline 
\noalign{\vskip\doublerulesep}
\textcolor{black}{$U|J=8|1$}  & \textcolor{black}{$-0.33$} & \textcolor{black}{$0.41$}\tabularnewline[\doublerulesep]
\noalign{\vskip\doublerulesep}
\textcolor{black}{$U|J=8|2$}  & \textcolor{black}{$-0.41$} & \textcolor{black}{$0.54$}\tabularnewline[\doublerulesep]
\noalign{\vskip\doublerulesep}
\textcolor{black}{$U|J=8|3$} & \textcolor{black}{$-0.52$} & \textcolor{black}{$0.65$}\tabularnewline[\doublerulesep]
\hline 
\end{tabular}
\end{table}

\textcolor{black}{As first examined by Kugel and Khomski\u{\i},\citep{Kugel'1982}
}\textcolor{black}{\emph{e}}\textcolor{black}{\textsubscript{%
\textcolor{black}{g}%
}\textsuperscript{%
\textcolor{black}{1}%
}} occupation polarization (\emph{i.e.}\textcolor{black}{, an electron-electron
Jahn-Teller degeneracy breaking) enhances virtual superexchange interactions,
relative to kinetic exchange interactions such as FM double-exchange.
This competition between superexchange and double-exchange is observed
in the LMO magnetic ground state, which varies according to the magnitude
of $\pi^{e_{\text{g}}}$ (}\textcolor{black}{\emph{e}}\textcolor{black}{\textsubscript{%
g %
} occupancy polarization). $\pi^{e_{\text{g}}}$ increases with $J$,
which explains the flip in long range magnetic ordering of the ground
state from FM to A-AFM as the intra-orbital parameter $J$ is increased.}

\textcolor{black}{At }\textcolor{black}{\emph{$J=2$}}\textcolor{black}{{}
eV the value of $\pi^{e_{\text{g}}}$ in Table }\ref{tab:Occupation-polarization-table}\textcolor{black}{{}
is large enough to stabilize the correct A-AFM ordering (as per Table
}\ref{table:UJrelaxed}\textcolor{black}{). The $U|J=8|2$ A-AFM ground
state in Figure \ref{fig:exptlmostruct} corresponds to a }$0.99(3z^{2}-r^{2})+0.41(x^{2}-y^{2})$
\textcolor{black}{\emph{e}}\textcolor{black}{\textsubscript{%
\textcolor{black}{g}%
} occupation density in the local octahedral basis. The orbital ordering
pattern across the unit cell is shown in Figure \ref{fig:Orbital-ordering-across-LMO},
and can be rationalized in terms of the Goodenough-Kanamori superexchange
rules.\citep{Goodenough1958,Kanamori1959} }

The $0.99(3z^{2}-r^{2})+0.41(x^{2}-y^{2})$ \textcolor{black}{\emph{e}}\textcolor{black}{\textsubscript{%
\textcolor{black}{g}%
} occupation density can be rewritten as }$1.97z^{2}-0.58x^{2}-1.4y^{2}$.
This expression shows the anisotropy in the \textcolor{black}{\emph{e}}\textcolor{black}{\textsubscript{%
\textcolor{black}{g}%
} state, in particular} between the $z$ and $x$ directions in the
octahedron: the $z^{2}$ contribution is much larger than $x^{2}$,
as per \textcolor{black}{Figure \ref{fig:Orbital-ordering-across-LMO}
b.} Each octahedral frame in the $ac$ plane is related to its neighbour
by a $\pi/2$ rotation about the $b$ lattice vector, so $z^{2}$/$x^{2}$
anisotropy forms a checker board pattern of \textcolor{black}{\emph{e}}\textcolor{black}{\textsubscript{%
\textcolor{black}{g}%
} partial occupation in the $ac$ plane. Note this corresponds to the
long/short Jahn-Teller Mn-O pattern }in the $ac$ plane\textcolor{black}{,
as per Figure \ref{fig:Orbital-ordering-across-LMO} c). According
to the Goodenough-Kanamori rules, superexchange in the $ac$ plane
is determined by }$z^{2}$/$x^{2}$ anisotropy in the\textcolor{black}{{}
}\textcolor{black}{\emph{e}}\textcolor{black}{\textsubscript{%
\textcolor{black}{g}%
} partial occupation}, and results in the FM coupling in the $ac$
plane.

The $y^{2}$ component of \textcolor{black}{the }\textcolor{black}{\emph{e}}\textcolor{black}{\textsubscript{%
\textcolor{black}{g}%
} partial occupation} forms occupied stripes pointed along local octahedra
y axes, following the \textcolor{black}{$b$ lattice direction (with
a small tilt) as in Figure \ref{fig:Orbital-ordering-across-LMO}
d). T}he continuous stripes of $y^{2}$ character along the $b$ lattice
direction correspond to the 'non-Jahn-Teller' Mn-O bonds in this direction.
The Goodenough-Kanamori rules determine that the continuous stripe
of $y^{2}$ character from the \textcolor{black}{\emph{e}}\textcolor{black}{\textsubscript{%
\textcolor{black}{g}%
} partial occupation }corresponds to AFM superexchange. The AFM coupling
is along the $b$ lattice parameter direction, between the FM coupled
$ac$ planes. \textcolor{black}{Together the in-plane FM and inter-plane
AFM produce the A-AFM ground state of LaMnO\textsubscript{%
\textcolor{black}{3}%
}, so our $U|J=8|2$ eV calculation results are in-line with experiment
as well. }

\textcolor{black}{If instead the Hund's coupling was weaker (}\textit{\textcolor{black}{i.e.}}\textcolor{black}{,
smaller }\textcolor{black}{\emph{J}}\textcolor{black}{), then $\pi^{e_{\text{\text{g}}}}$
would also be smaller. This alters the character of the} occupied
states in the \emph{e}\textsubscript{%
g%
} shell,\textcolor{black}{{} so that orbital-ordering mediated A-AFM
superexchange is reduced relative to other effects such as FM double-exchange.
This explains why }stabilization of A-AFM magnetic ordering (see\textcolor{black}{{}
Table }\ref{table:UJrelaxed}) is only possible when intra-orbital
exchange is large enough\textcolor{black}{. Too small of an intra-orbital
exchange interaction is the origin of the incorrect FM ground state
found in prior examinations\citep{Sawada1997,Solovyev1996,Hashimoto2010}
of LMO using standard DFT. }

\textcolor{black}{The improvements in the LMO description through
applying exchange corrections reinforce hints by }Sawada \emph{et
al}.,\textcolor{black}{\citep{Sawada1997}} Solovyev \emph{et al.}\textcolor{black}{\citep{Solovyev1996}}
and Hashimoto \emph{et al}\textcolor{black}{.,\citep{Hashimoto2010}
that the correct orbital and magnetic ordering in LMO requires an
anisotropic intra-orbital exchange correction to the DFT ground state.}
In what follows, we discuss further details of the electronic and
crystal structure.

\subsection{Electronic and crystal structure details}

\subsubsection{Orbital order}

\begin{figure}
\includegraphics[scale=0.1]{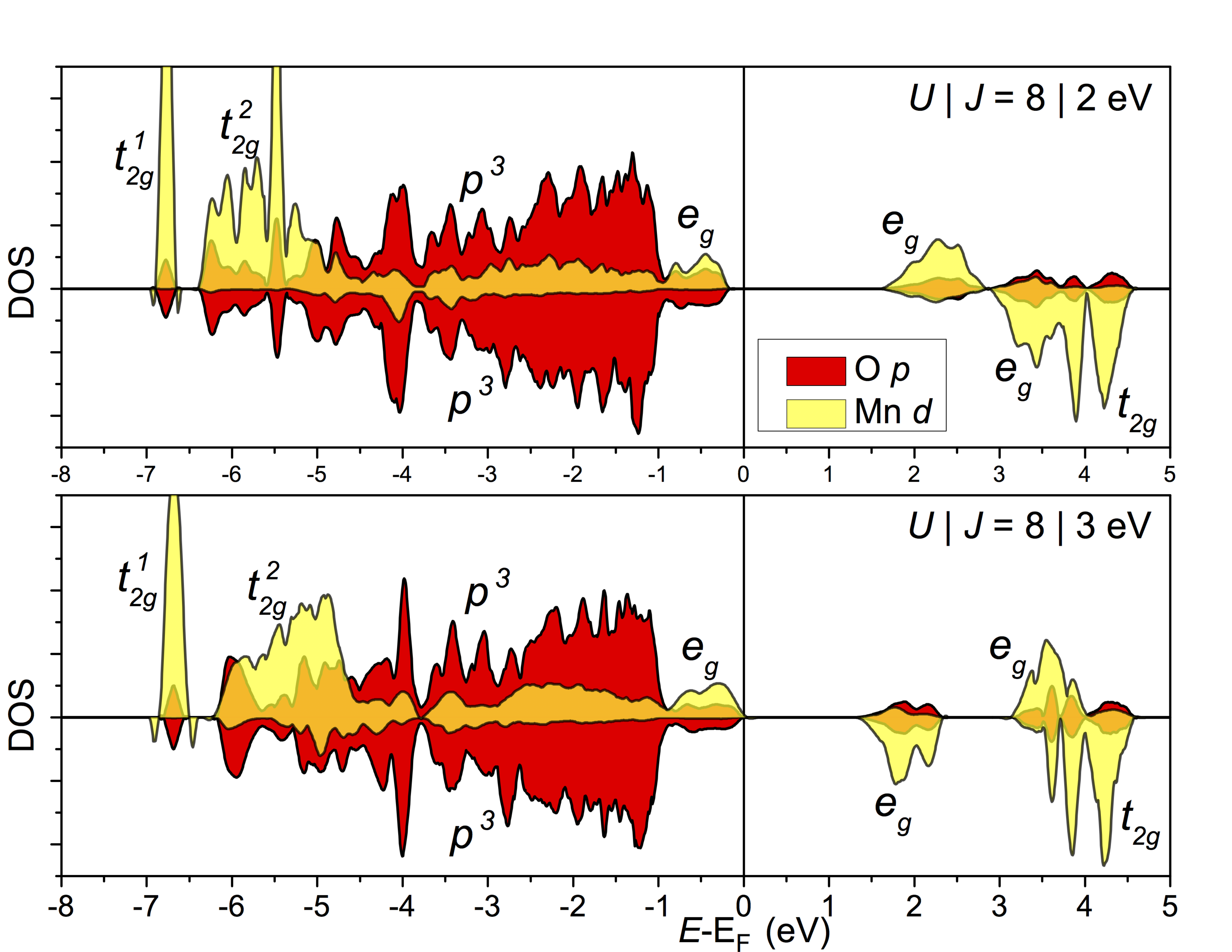}

\caption{%
\label{fig:UJ_82_83_DOS_detailed}LaMnO\textsubscript{3} densities
of states (DOS) for $U|J=8|2$ eV and $U|J=8|3$ eV calculations.
Majority spin corresponds to positive DOS and minority to negative
DOS.%
}
\end{figure}

\begin{figure}
\textcolor{black}{\includegraphics[scale=0.32]{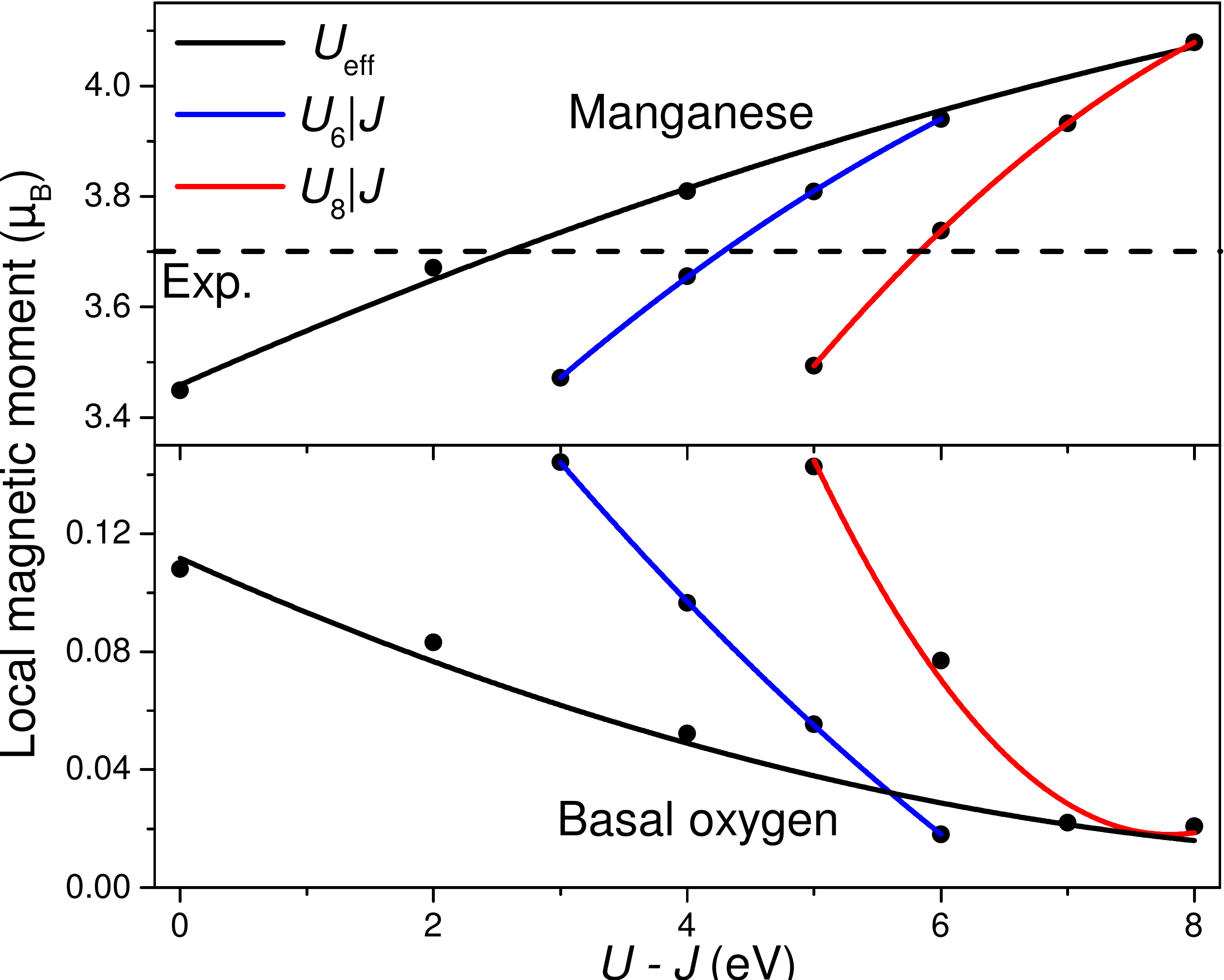}}

\textcolor{black}{\caption{%
\textcolor{black}{Local magnetic moment within the Bader volume for
Mn cations and O anions (within the ($010$) basal plane), from DFT+$U_{\text{eff}}$
and DFT+}\textcolor{black}{\emph{$U|J$}}\textcolor{black}{{} methods}\textcolor{black}{\emph{.}}\textcolor{black}{{}
The notation}\textcolor{black}{\emph{ $U_{6}|J$ }}\textcolor{black}{and
$U_{8}|J$ indicates $U$ is fixed to 6 eV and 8 eV, respectively,
while $J$ is varied.}\textcolor{black}{\emph{ }}\textcolor{black}{The
experimental reference local magnetic moment is $3.7$}\textcolor{black}{\emph{
$\mu_{\text{B}}$.}}\textcolor{black}{\citep{Elemans1971}}\textcolor{green}{\emph{
}}\label{fig:Bade-plt-1}%
}
}
\end{figure}

It was previously shown that applying Coulomb corrections, such as
with $U|J=8|2$ eV, corrected the LMO electronic, magnetic and lattice
structure. Further electronic structure details are shown for the
LMO DOS in Figure \ref{fig:UJ_82_83_DOS_detailed} at the $U|J=8|2$
eV level of correction. In Figure \ref{fig:UJ_82_83_DOS_detailed}
the position of each band in the Mn DOS agree quantitatively with
the optical conductivity measurements of Jung \emph{et al}..\citep{Jung1997}
Further experimental agreement comes from our $U|J=8|2$ eV calculated
local magnetic moment, which at 3.7$\mu_{\text{B}}$ agrees with Eleman's
measurement.\citep{Elemans1971} The $U$ and $J$ dependence of the
local magnetic moments are shown in Figure \ref{fig:Bade-plt-1}.
The high sensitivity of the electronic structure of LMO to perturbations
in part underlies its complex phase diagram. This is illustrated by
comparing the ${\color{black}U|J=8|2}$ and ${\color{black}U|J=8|3}$
eV DOS in Figure \ref{fig:UJ_82_83_DOS_detailed}, and examining the
magnetic state of DOS near the edges of the valence band maximum (VBM)
and conduction band minimum (CBM). For ${\color{black}U|J=8|2}$ eV,
Hund's rules are obeyed as both VBM and CBM have the same spin state
whereas Hund's rules are broken for ${\color{black}U|J=8|3}$ eV.
We find that the cross-over occurs at $J\approx2.4$ eV. LaMnO\textsubscript{%
3%
} is fragile in terms of exchange: above $J\approx2.4$ eV we have
the breakdown of Hund's rules while below $J\approx1.8$ eV incorrectly
stabilizes the FM rather than A-AFM ground state. 

\textcolor{black}{}

\subsubsection{Magnetic coupling constants}

The magnetic coupling constants in LMO have been extracted by Mu$\tilde{\text{n}}$oz
\emph{et al.}\citep{Munoz2004} amongst others\citep{Lee2013,Hashimoto2010,Nicastro2002,Solovyev1996},
by considering an Ising model (with $S=2$ spin moment per Mn ion)
for different spin-ordered solutions. The intra-plane ($ac$) $J_{1}$
and inter-plane ($b$) $J_{2}$ coupling constants for the $20$ atom
LMO unit cell are

\begin{eqnarray}
J_{1} & = & \frac{1}{64}\left[E^{\text{G-AFM}}-E^{\text{A-AFM}}\right]\nonumber \\
J_{2} & = & \frac{1}{32}\left[E^{\text{A-AFM}}-E^{\text{FM}}\right]\,.\label{eq: Exchange couplings J1 J2}
\end{eqnarray}

The initial A-AFM/FM stability results in Table \ref{table:UJrelaxed}
hint that the coupling constants will depend strongly on the Hund's
exchange parameter. In the context of previous works, $J_{i}$ are
well known to be highly sensitive, for example to variation in Mn-O-Mn
angle through superexchange interactions\citep{Meskine2001}, and
the Mn ionic charge population\citep{Nicastro2002}.

\begin{figure}
\centering{}\includegraphics[scale=0.4]{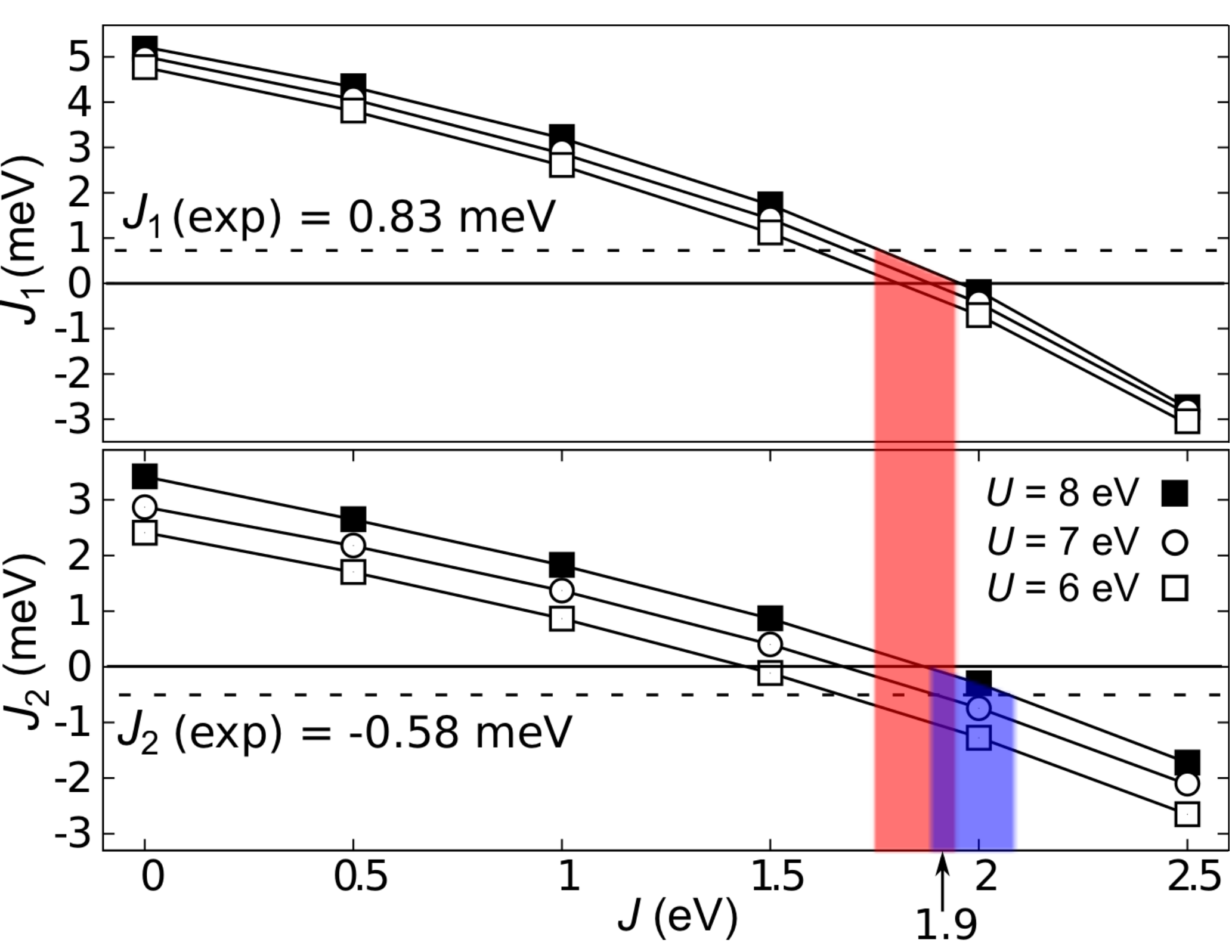}\caption{%
LaMnO\textsubscript{3} magnetic coupling constants $J_{1}$ and $J_{2}$
versus $U|J$ schema Hund's exchange parameter $J$, for $U=6$ eV
(white squares), $U=7$ eV (white circles), $U=8$ eV (black squares).
$J_{1}$ and $J_{2}$ are defined in Equation (\ref{eq: Exchange couplings J1 J2}).
The red-blue overlap (centered at $J=1.9$ eV) suggests a $J$ exchange
value for the $U|J$ scheme that provides the correct sign for both
coupling constants - see main text for discussion.\label{fig:LaMnO-magnetic-coupling}%
}
\end{figure}

On the trend of magnetic stability in $U$ and $J$, the superexchange
interaction which stabilizes AFM ordering is expected to scale inversely
with effective on-site Coulomb interaction, \emph{i.e.} $\sim t^{2}/U_{\text{eff}}$
where $t$ is effective inter-site hopping. Considering first $J_{2}$
($\sim E^{\text{A-AFM}}-E^{\text{FM}}$) in Figure \ref{fig:LaMnO-magnetic-coupling},
the stability of AFM coupling along $b$ decreases both with increasing
$U$ or decreasing $J$ as expected since $U_{\text{eff}}=U-J$. However,
the dependence of the $J_{2}$ coupling on $U$ and $J$ is not equivalent:
the variation in $J_{2}$ is some five-fold more sensitive to changes
in $J$ than $U$, \emph{i.e.}, $\partial J_{2}/\partial J\,\approx-5\partial J_{2}/\partial U\,$.
The AFM coupling in the $ac$ plane, measured by $J_{1}$, is even
more sensitive to the intra-orbital Hund's parameter, with $\partial J_{1}/\partial J\,\approx-10\partial J_{1}/\partial U\,$.
The origin of coupling constant sensitivity to $J$, is the strongly
anisotropic effect of $J$ on the LMO Mn \emph{d} shell states, with
variation in $J$ increasing $\pi^{e_{g}}$ in Table \ref{tab:Occupation-polarization-table}
above and beyond that accesible with $U$ alone.

Neutron scattering experiments have determined coupling constant values
of $J_{1}^{\text{exp}}=0.83$ meV and $J_{2}^{\text{exp}}=-0.58$
meV.\citep{Rodriguez-Carvajal1998} In Figure \ref{fig:LaMnO-magnetic-coupling}
reasonable values for $J_{1}$ are produced with $J\approx1.75$ eV,
and for $J_{2}$ with $J_{2}\approx2$ eV. The discrepancy in $J$
value for each $J_{i}$ is perhaps unsurprising given the extreme
sensitivity of A-AFM, G-AFM and FM phases to the intra-orbital Hund's
interaction. Overall the $U|J=8|2$ eV combination previously suggested
remains a good compromise at the level of half integer eV screening
intervals considered here. Although higher resolution screening in
$J$ is beyond the scope of this work, if DFT+$U|J$ calculations
are required for thermodynamics applications, results indicate a small
modification of $J$ by a few percent may be advantageous to tune
the magnetic transition temperatures precisely, while the magnetic
couplings are relatively insensitive to the direct $U$ term.

Due to the sensitivity of the magnetic couplings to the Coulombic
$J$ correction, agreement with experiment is challenging. At the
$U=8$ eV required to open the band gap, and screening in $J$ at
half integer intervals shown in Figure \ref{fig:LaMnO-magnetic-coupling},
$U|J=8|2$ eV remains the best compromise. For $U|J=8|2$ eV the inter-plane
coupling at $J_{2}=-0.30$ meV has the correct sign but in magnitude
falls short of $J_{2}^{\text{exp}}=-0.58$ meV\citep{Rodriguez-Carvajal1998}.
More problematic is the intra-plane coupling, which overestimates
the tendency for electrons to couple antiferromagnetically in the
$ac$ plane, excessively stabilizing G-AFM ordering at $J_{2}=-0.19$
meV compared to $J_{2}^{\text{exp}}=+0.83$ meV. 

In Figure \ref{fig:LaMnO-magnetic-coupling} the colored areas show
the $J$ values that correspond to coupling constants between zero
and $J_{i}^{\text{exp}}$, \emph{i.e.} the correct sign for each $J_{i}$.
The overlap in colored areas identifies the narrow range of intra-orbital
exchange values, $1.88\le J\le1.95$ eV, that gives the correct signs
for both $J_{i}$ together, with $E(\text{A-AFM)}<E(\text{FM)}<E(\text{G-AFM)}$
in agreement with experiment\citep{Rodriguez-Carvajal1998}. Based
on the refinement in $J$ value, we have performed GGA (PBEsol) calculations
with $U|J=8|1.9$ eV. The error in calculation results with respect
to experimental values\citep{Jung1997,Jung1998,Rodriguez-Carvajal1998,SakaiN.;FjellvagH.;Lebech1997,Elemans1971,Moussa1996}
is summarized in Table \ref{tab: j1.9 LMO properties}. The $U|J=8|1.9$
eV calculations produce good experimental agreement overall, with
magnetic coupling constants with signs that agree with experiment,
$J_{1}=+0.2$ and $J_{2}=-0.1$ meV, a band gap value only a couple
of percent above the experimental 1.7 eV value, lattice parameter
errors between $+1.5$ \% and $-0.8$ \% which largely cancel to give
a volume error with respect to experiment of $+0.1$ \%.\citep{Rodriguez-Carvajal1998,Jung1997}

\begin{table*}
\caption{%
LaMnO\textsubscript{3} electronic, magnetic and structural properties
obtained from a $U|J=8|1.9$ eV calculation, with comparison to experimental
counter-parts.\citep{Jung1997,Jung1998,Rodriguez-Carvajal1998,SakaiN.;FjellvagH.;Lebech1997,Elemans1971,Moussa1996}
The $J=1.9$ eV value is based on a refinement of the Hund's exchange
parameter to secure the correct sign for both magnetic coupling constants,
$J_{1}$ and $J_{2}$, which are exceptionally sensitive to on-site
exchange - see Figure \ref{fig:LaMnO-magnetic-coupling}. \label{tab: j1.9 LMO properties}%
}

\centering{}%
\begin{tabular}{cccccccc}
\hline 
\multirow{2}{*}{$U|J$ (eV)} & \multicolumn{2}{c}{Electronic gap} & \multicolumn{2}{c}{Magnetic} & \multicolumn{3}{c}{Structural}\tabularnewline
 & $E^{\text{\,\ Gap}}$ (eV) \foreignlanguage{english}{\citep{Jung1997}} & Character\foreignlanguage{english}{ \citep{Jung1997}} & $J_{1}$, $J_{2}$ (meV) \foreignlanguage{english}{\citep{Moussa1996}}  & $M$ ($\mu_{\text{B}}$) \foreignlanguage{english}{\citep{Elemans1971}} & $\mathbf{Q}^{\text{Ortho}}$, $\mathbf{Q}^{\text{Tetra}}$ (a.u.)
\foreignlanguage{english}{\citep{SakaiN.;FjellvagH.;Lebech1997}} & $a$, $b$, $c$ (${\rm \AA}$) \foreignlanguage{english}{\citep{SakaiN.;FjellvagH.;Lebech1997}} & $V$ (${\rm \AA}^{3}$) \foreignlanguage{english}{\citep{SakaiN.;FjellvagH.;Lebech1997}}\tabularnewline
\hline 
\hline 
\noalign{\vskip0.1cm}
$8|1.9$ & $1.75$  & $e_{g\uparrow}^{1}\to e_{g\uparrow}^{2}$  & $+0.2,\,-0.1\,\,$  & $3.76$  & $0.145$, $0.856\,\,$  & $5.823,\,7.642,\,5.508\,$  & $245$ \tabularnewline[0.1cm]
\noalign{\vskip0.1cm}
\emph{Exp. }%
 & %
$1.7\,$%
 & $e_{g\uparrow}^{1}\to e_{g\uparrow}^{2}\,$  & %
$+0.83$, $-0.58\,$%
 & %
$3.7\,$%
 & %
$0.14$, $0.78\,$%
 & %
$5.736,\,7.703,\,5.540\,$%
 & %
$245\,$%
\tabularnewline[0.1cm]
\hline 
\end{tabular}
\end{table*}

\subsubsection{Jahn-Teller distortion}

\begin{figure*}
\begin{centering}
\includegraphics[scale=0.66]{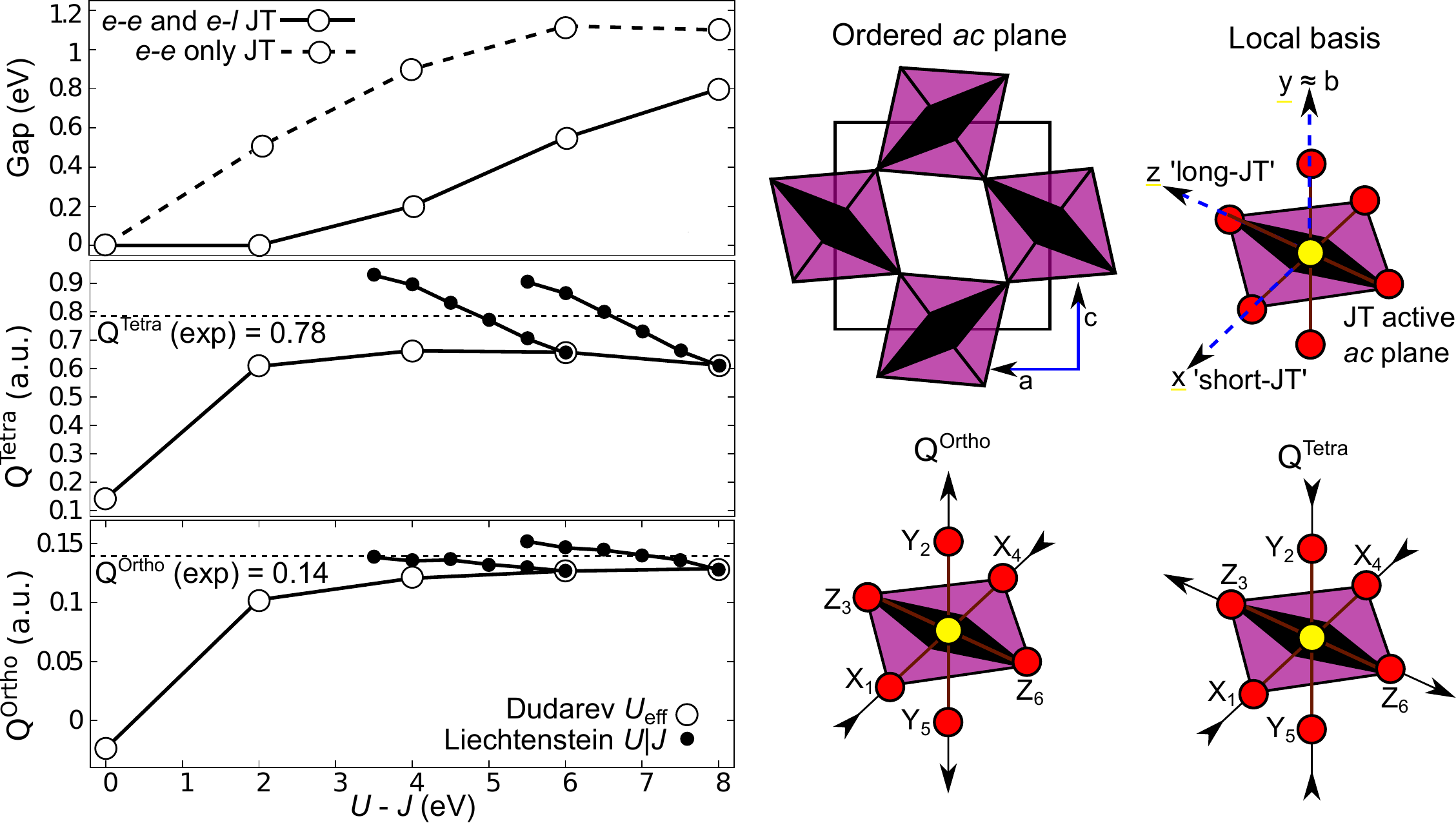}
\par\end{centering}

\caption{%
\emph{Left top panel:} Band gap of LaMnO\textsubscript{3} versus
$U-J$ within the $U_{\text{eff}}$ approach for fully relaxed structures
where both electron-lattice and electron-electron interactions are
active (dashed line, \textit{e-l} and \textit{e-e}) and for structures
with the Jahn-Teller distortion frozen out so only electron-electron
interactions are active (solid line, \textit{e-e} only). \emph{Left
bottom and middle panels:} Jahn-Teller normal modes versus $U-J$
within the $U_{\text{eff}}$ approach (white circles), and within
the \emph{$U|J$} approach (black circles) for $U$ fixed to $8$
eV and $6$ eV with $J$ varied. \emph{Right panels:} Orbitally ordered
and strongly Jahn-Teller active \emph{ac} plane, the local basis convention,
and $\mathbf{Q}^{\text{Ortho}}$ and $\mathbf{Q}^{\text{Tetra}}$
modes. \label{fig:Jahn-Teller-gap}%
}
\end{figure*}

We end our analysis with conclusions on the nature of Jahn-Teller
distortion in LMO and on the origin of the LMO insulating state. Jahn-Teller
distortion in LaMnO\textsubscript{%
3%
} can be characterized in terms of two normal modes of the type introduced
by van Vleck\citep{VanVleck1939} and by Kanamori\citep{Kanamori1960}.
The normal modes are shown in Figure \ref{fig:Jahn-Teller-gap} along
with the crystal unit cell and the local octahedral basis. The modes
are calculated as

\begin{eqnarray*}
\mathbf{Q}^{\text{Ortho}} & = & \frac{1}{\sqrt{2}}\left[\mathbf{Y}_{2}-\mathbf{Y}_{5}-\mathbf{X}_{1}+\mathbf{X}_{4}\right]\\
\mathbf{Q}^{\text{Tetra}} & = & \frac{1}{\sqrt{6}}\left[2\mathbf{Z}_{3}-2\mathbf{Z}_{6}-\mathbf{Y}_{2}+\mathbf{Y}_{5}-\mathbf{X}_{1}+\mathbf{X}_{4}\right]\,.
\end{eqnarray*}
Each variable represents an octahedral bond length, with subscripts
indexing oxygen octahedral cage sites for a given manganese centre,
$i\,(i=1,\,...6)$. In the local basis in this work, which differs
from other choices\citep{Hashimoto2010,Lee2013,Khomskii2014}, $\mathbf{Z}_{i}=\mathbf{z}_{i}^{\text{O}}-\mathbf{z}^{\text{Mn}}$
are the long Mn-O bonds and $\mathbf{X}_{i}=\mathbf{x}_{i}^{\text{O}}-\mathbf{x}^{\text{Mn}}$
short Mn-O bonds, with both in the FM coupled \emph{ac} plane\emph{.
}$\mathbf{Y}_{i}=\mathbf{y}_{i}^{\text{O}}-\mathbf{y}^{\text{Mn}}$
are along the inter-plane AFM coupled \emph{b} lattice direction. 

We begin commenting that the formation of a band gap in LaMnO\textsubscript{%
3%
} is not solely electron-electron (\emph{e-e}) or electron-lattice
(\emph{e-l}\textit{\emph{)}} in character. Rather, it is a joint function
of the lattice relaxation and development of Jahn-Teller distortions
as well as the strong on-site Coulomb interaction. This is illustrated
explicitly in Figure \ref{fig:Jahn-Teller-gap}. As mentioned above,
two logically distinct routes to breaking symmetry exist in order
to produce a gap: (i) a purely electronic effect \emph{via} electron-electron
interactions and the formation of a sizable orbital polarization $\pi^{e_{\text{g}}}$
that breaks symmetry (also called \textit{e-e} Jahn-Teller distortion)\citep{Kugel'1982},
or (ii) electron-lattice (\textit{e-l}) Jahn-Teller distortions where
certain local octahedral phonon modes become soft, the Mn-O bond lengths
become unequal, and this creates crystal field symmetry breaking.
These two mechanisms are in fact mutually enhancing, and which one
causes which in LaMnO\textsubscript{%
3%
} is an open question that has been debated in the works of Khomski\u{\i}\citep{Khomskii2005},
Yin \emph{et al}.\citep{Yin2005}, and Loa \emph{et al}.\citep{Loa2001}. 

In some materials, one mechanism can clearly dominate over the other.
For example, in KCuF\textsubscript{%
3%
}, to which LaMnO\textsubscript{%
3%
} is superficially similar as both are perovskites with partial $e_{\text{g}}$
occupation, the symmetry lowering is truly driven by electronic interactions
alone,\citep{Liechtenstein1995a} and thus KCuF\textsubscript{%
3%
} is said to exhibit \emph{e-e} Jahn-Teller distortion. Figure \ref{fig:Jahn-Teller-gap}
shows that the nature of Jahn-Teller is different in LaMnO\textsubscript{%
3%
}. 

Firstly, with the \textit{e-l} distortion frozen out, one can generate
symmetry breaking and a gap for a Coulomb interaction strength \textit{\emph{(}}\textit{$U-J$}\textit{\emph{)}}
above a critical value $\sim2$ eV, so that in principle, the lattice
distortion is not necessary to create a gap. However, in practice,
the gap and orbital splitting remain small without lattice Jahn-Teller
distortions. Secondly, with \textit{$U-J$} set to zero, the DFT calculations
do produce weakly active \textit{e-l} Jahn-Teller distortions of $\mathbf{Q}^{\text{Ortho}}=-0.02$
a.u. and $\mathbf{Q}^{\text{Tetra}}=0.14$ a.u., but the gap remains
essentially zero. The addition of Coulomb repulsion \emph{via} \textit{\emph{$U_{\text{eff}}$}}
greatly enhances the \textit{e-l} distortion of each mode to approximately
$\mathbf{Q}^{\text{Ortho}}\approx0.12$ a.u. and $\mathbf{Q}^{\text{Tetra}}\approx0.62$
a.u.. However even with $U_{\text{eff}}$ applied $\mathbf{Q}^{\text{Ortho}}$
and $\mathbf{Q}^{\text{Tetra}}$ remain still short of experiment
by some $13$ \% and $20$ \% respectively. As per Table \ref{tab:Occupation-polarization-table}
and Figure \ref{fig:Jahn-Teller-gap}, one can only go so far with
$U_{\text{eff}}$: the orbital polarization $\pi^{e_{g}}$ is too
weak and the Jahn-Teller \textit{e-l} distortion remains largely unchanged
with increasing $U_{\text{eff}}$. 

The only way to bridge the deficit is through the use of a dedicated
exchange term \textit{via} the $U|J$ approach. As shown in Table
\ref{tab:Occupation-polarization-table}, \emph{J} increases $\pi^{e_{\text{g}}}$
and anisotropy throughout the \emph{d} manifold significantly. By
increasing $J$ in the $U|J$ scheme, the LMO $\mathbf{Q}^{\text{Ortho}}$
and $\mathbf{Q}^{\text{Tetra}}$ modes can be tuned to agree with
experiment by accessing additional \textit{e-e} Jahn-Teller activity
otherwise unavailable.

\section{Conclusion}

An isotropic Hubbard correction, such as the $U_{\text{eff}}$ methodology,
is unable to simultaneously reproduce the band gap, experimental level
of Jahn-Teller distortion and magnetic ordering of bulk LaMnO\textsubscript{%
3%
}. At small $U_{\text{eff}}$\emph{, }A-AFM magnetic ordering is correctly
stabilized but the gap and structural distortions are underestimated.
With increasing $U_{\text{eff}}$ values, the gap and crystal structure
are reproduced but FM ordering is incorrectly stabilized. The \emph{$U|J$}
approach, with its explicit exchange dependence on orbital symmetry,
provides a better picture of electronic, magnetic and structural properties
of LaMnO$_{3}$. The origin of the \emph{$U|J$} success is the Hund's
coupling accounted for by the spatial/orbital dependence of the dedicated
exchange terms that depend on \emph{J. }\textit{\emph{These terms}}\emph{
}selectively polarize orbital occupation through highly anisotropic
energy splitting in the Mn \textit{d} manifold. Only the addition
of \emph{J} terms, rather than crystal field or direct Coulomb \emph{U},
can provide appropriate and large enough anisotropic splitting within
the $t_{\text{2g}}$ and $e_{\text{g}}$ manifolds. Orbital order
due to the short range \emph{J} makes possible the combination of
long-range FM exchange in the $(010)$ plane, and AFM exchange between
\{$010$\} planes, to stabilize the A-AFM ordered ground state. Soft
phonon modes (\emph{e-l} Jahn-Teller) and electronic occupation polarization
(\emph{e-e} Jahn-Teller) contribute jointly to the insulating state,
with the latter predominant. The experimental Jahn-Teller distortion
magnitude can only be achieved by adding the anisotropy \emph{J} provides
on top of the direct Coulomb occupancy polarization. The best description
of LaMnO\textsubscript{3} is achieved within the PBEsol+\emph{U}
framework when $U=8$ eV and $J=1.9$ eV.

\section{References}

\bibliographystyle{rsc}
\bibliography{LaMnO3_resubmission}

\begin{acknowledgments}
This collaborative work was funded by grants from the US\textquoteright{}s
National Science Foundation (NSF-DMR MRSEC 1119826) and from the UK\textquoteright{}s
Engineering and Physical Sciences Research Council EPSRC (EP/J001775/1).
Via the UK\textquoteright{}s HPC Materials Chemistry Consortium, which
is funded by EPSRC (EP/L000202), this work made use of HECToR and
ARCHER, the UK\textquoteright{}s national high-performance computing
services.
\end{acknowledgments}
\rule[0.5ex]{1\columnwidth}{1pt}

\appendix

\section*{Appendix A: DFT+\emph{U} expressions\label{sec:Appendix-A}}

We begin with the $U|J$ rotationally invariant DFT+\emph{U} total
energy expression\citep{Liechtenstein1995a} written for a single
atomic site (since the corrections are linear sums over atomic sites),

\[
E_{\text{DFT+}U|J}=E_{\text{DFT}}+E_{U}-E_{\text{dc}}\,\,.
\]
$E_{\text{DFT}}$ is the total DFT energy using some flavor of exchange
and correlation, the Coulombic $+U$ correction energy is

\begin{multline*}
E_{U}=\frac{1}{2}\sum_{\sigma,\sigma',m^{i}}(m\sigma m''\sigma'|V|m'\sigma m'''\sigma')\times\\
(\rho_{m'm}^{\sigma}\rho_{m'''m''}^{\sigma'}-\rho_{m'''m}^{\sigma}\rho_{m'm''}^{\sigma}\delta_{\sigma\sigma'})
\end{multline*}
and the double-counting correction $E_{\text{dc}}$ is

\[
E_{\text{dc}}=\sum_{\sigma}\frac{(U-J)}{2}N_{\sigma}(N_{\sigma}-1)+\frac{{U}}{2}N_{\sigma}N_{\bar{\sigma}}\,\,.
\]

In the above expressions, $V(\mathbf{r},\mathbf{r}')=1/|\mathbf{r}-\mathbf{r}'|$
is the bare Coulomb interaction, $\sigma$ labels spin where $\bar{\sigma}$
is the opposite spin to $\sigma$, $m$ labels angular momentum states
of the atomic shell under consideration (\emph{d} orbitals for Mn
in this paper), $\rho_{mm'}^{\sigma}$ is the single-particle density
matrix, $N_{\sigma}=trace(\rho^{\sigma})=\sum_{m,m'}\rho_{mm'}^{\sigma}\delta_{mm'}$
is the number of electrons on the site of spin $\sigma$, and $U$
and $J$ are the direct and exchange Coulomb parameters. 

The matrix elements of $V$ are defined by $(m\sigma m''\sigma'|V|m'\sigma m'''\sigma')=\intop dr\intop dr'\,\,\phi_{m\sigma}^{*}(\mathbf{r})\phi_{m'\sigma}(\mathbf{r})\,\frac{1}{|\mathbf{r}-\mathbf{r}'|}\,\phi_{m''\sigma'}^{*}(\mathbf{r'})\phi_{m'''\sigma'}(\mathbf{r}')$.
The matrix elements of $V$ are further decomposed for an atomic shell
with angular momentum $l$ \emph{via}

\begin{multline*}
(m\sigma m''\sigma'|V|m'\sigma m'''\sigma')=\\
\delta_{m-m',m'''-m''}\sum_{k=0}^{2l}c^{k}(lm,lm')c^{k}(lm''',lm'')F^{k}
\end{multline*}
where $c^{k}$ and $F^{k}$ are standard atomic Slater angular integrals
and radial integrals. For \emph{d} shells, $U=F^{0}$, $J=(F^{2}+F^{4})/14$
and $F^{4}/F^{2}=0.625$ are the canonical choices\citep{Liechtenstein1995a}.
Thus only two parameters are needed to specify the radial integrals:
$F^{0}=U$, $F^{2}=(112/13)J$ and $F^{4}=(70/13)J$. 

To make progress with expressions for $E_{U}$ and $E_{\text{dc}}$
which are given in terms of $\rho_{mm'}^{\sigma}$ and $N_{\sigma}$,
we need rewrite these expressions in terms of the occupancy eigenvalues
of the density matrix, $f_{i\sigma}$. Denoting the eigenvectors of
$\rho_{mm'}^{\sigma}$ as $\text{V}_{m,i}^{\,\sigma}$ so that
\[
\rho_{mm'}^{\sigma}=\sum_{i}\text{V}_{m,i}^{\,\sigma}\,\, f_{i\sigma}\,\left(\text{V}_{m',i}^{\,\sigma}\right)^{*}
\]
we may insert this expansion into the expression for $E_{U}$. After
some algebraic manipulations, using the fact that $c^{0}(lm,lm')=\delta_{mm'}$
for the $k=0$ term and the unitarity of the eigenvector $\text{V}^{\,\sigma}$
matrices, we find
\begin{multline*}
E_{U}=\frac{{U}}{2}\left(N^{2}-\sum_{i\sigma}f_{i\sigma}^{2}\right)+\\
\frac{{1}}{2}\sum_{\sigma,\sigma',i,j}\text{C}_{ij}^{\,\sigma\sigma'}f_{i\sigma}f_{j\sigma'}-\text{X}_{ij}^{\sigma}f_{i\sigma}f_{j\sigma}\delta_{\sigma\sigma'}
\end{multline*}
where $N=\sum_{\sigma}N_{\sigma}$ is the total electron count on
the site, and the Coulombic\textcolor{red}{{} }\textcolor{black}{$\text{C}^{\,\sigma\sigma'}$
and exchange $\text{X}^{\,\sigma}$} correction matrices are given
by
\begin{multline*}
\text{C}_{ij}^{\,\sigma\sigma'}=\sum_{k=2}^{2l}F^{k}\sum_{mm'm''m'''}\delta_{m-m',m'''-m''}\times\\
(\text{V}^{\,\sigma})_{im}^{\dag}c^{k}(lm,lm')\text{V}_{m'i}^{\,\sigma}(\text{V}^{\,\sigma'})_{jm'''}^{\dag}c^{k}(lm''',lm'')\text{V}_{m''j}^{\,\sigma'}
\end{multline*}
and
\begin{multline*}
\text{X}_{ij}^{\sigma}=\sum_{k=2}^{2l}F^{k}\sum_{mm'm''m'''}\delta_{m-m',m'''-m''}\times\\
(\text{V}^{\,\sigma})_{im}^{\dag}c^{k}(lm,lm')\text{V}_{m'j}^{\,\sigma}\,\,(\text{V}^{\,\sigma})_{im'''}^{\dag}c^{k}(lm''',lm'')\text{V}_{m''j}^{\,\sigma}\,.
\end{multline*}
The Coulomb correction $\text{C}^{\,\sigma\sigma'}$ matrices have
zero average over all entries, a fact easily shown by using some basic
properties of the Slater angular integrals. The same can be done for
the exchange correction matrices by separating out a constant term

\[
\text{X}_{ij}^{\,\sigma}=\Delta\text{X}_{ij}^{\,\sigma}+J(1-\delta_{ij})\,.
\]
Substituting this into the previous $E_{U}$ expression and subtracting
the double-counting term $E_{\text{dc}}$ to cancel common terms then
yields the total energy 
\begin{multline*}
E_{\text{DFT}+U|J}=E_{\text{DFT}}+\frac{U-J}{2}\sum_{i\sigma}(f_{i\sigma}-f_{i\sigma}^{2})+\\
\frac{1}{2}\sum_{\sigma,\sigma',i,j}\text{C}_{ij}^{\,\sigma\sigma'}f_{i\sigma}f_{j\sigma'}-\Delta\text{X}_{ij}^{\,\sigma}f_{i\sigma}f_{j\sigma}\delta_{\sigma\sigma'}
\end{multline*}
which is in the form of the DFT+$U_{\text{eff}}$ (Dudarev) energy\citep{Dudarev1998}
plus a correction involving the $\text{C}^{\,\sigma\sigma'}$ and
$\Delta\text{X}^{\,\sigma}$ matrices and the occupancies. Therefore,
the $U|J$ scheme can be viewed as a correction to the $U_{\text{eff}}$
method which includes additional Coulombic and exchange terms stemming
from exchange integrals between different orbitals: this is because
both $\text{C}^{\,\sigma\sigma'}$ and $\Delta\text{X}^{\,\sigma}$
are proportional to $J$ and thus the orbital shape dependence of
the Coulombic interactions on the site, something neglected by the
$U_{\text{eff}}$ scheme. 

The correction to the eigenvalue follows from the occupancy derivative
of the added terms to the DFT energy
\begin{multline*}
\frac{\partial(E_{U}-E_{\text{dc}})}{\partial f_{i\sigma}}=(U-J)\left(\frac{1}{2}-f_{i\sigma}\right)+\\
\sum_{j\sigma'}\text{\text{C}}_{ij}^{\,\sigma\sigma'}f_{j\sigma'}-\Delta\text{X}_{ij}^{\,\sigma}f_{j\sigma}\delta_{\sigma\sigma'}\,.
\end{multline*}
In what follows, it is more convenient to work with vectors and matrices.
Thus if we collect all occupancies $f_{i\sigma}$ into a column vector
$f_{\sigma}$, then the above eigenvalue correction can be more compactly
written as
\begin{multline*}
\nabla_{f_{\sigma}}(E_{U}-E_{\text{dc}})=(U-J)\left(\frac{1}{2}-f_{\sigma}\right)+\\
J\left[\text{A}^{\sigma}f_{\sigma}+\text{B}^{\sigma}f_{\bar{\sigma}}\right]
\end{multline*}
where we have peeled off the constant $J$ and also indicated same
spin and opposite spin occupancy dependences \emph{via} the unitless
matrices 
\[
\text{A}^{\sigma}=(\text{C}^{\,\sigma\sigma}-\Delta\text{X}^{\,\sigma})/J
\]
and 
\[
\text{B}^{\sigma}=\text{C}^{\sigma\bar{\sigma}}/J\,.
\]

We now proceed to actual example cases to compute numerical values
for the $\text{A}^{\sigma}$ and $\text{B}^{\sigma}$ matrices. The
simplest assumption is to take the spherical harmonic states $Y_{lm}$
as the eigenbasis of the density matrix $\rho^{\sigma}.$ This means
$\text{V}^{\sigma}=\text{I}$ and one can directly compute the matrices
using numerical values for Slater angular integrals. The results are
\[
\text{A}^{\sigma}=\left(\begin{array}{r|rrrrr}
 & Y_{22} & Y_{21} & Y_{20} & Y_{2,-1} & Y_{2,-2}\\
\hline Y_{22} & 0 & -0.52 & -0.52 & 0.17 & 0.86\\
Y_{21} & -0.52 & 0 & 0.52 & -0.17 & 0.17\\
Y_{20} & -0.52 & 0.52 & 0 & 0.52 & -0.52\\
Y_{2,-1} & 0.17 & -0.17 & 0.52 & 0 & -0.52\\
Y_{2,-2} & 0.86 & 0.17 & -0.52 & -0.52 & 0
\end{array}\right)\,
\]

and
\[
\text{B}^{\sigma}=\left(\begin{array}{r|rrrrr}
 & Y_{22} & Y_{21} & Y_{20} & Y_{2,-1} & Y_{2,-2}\\
\hline Y_{22} & 0.72 & -0.40 & -0.63 & -0.40 & 0.72\\
Y_{21} & -0.40 & 0.37 & 0.06 & 0.37 & -0.40\\
Y_{20} & -0.63 & 0.06 & 1.14 & 0.06 & -0.63\\
Y_{2,-1} & -0.40 & 0.37 & 0.06 & 0.37 & -0.40\\
Y_{2,-2} & 0.72 & -0.40 & -0.63 & -0.40 & 0.72
\end{array}\right)\,.
\]
However, this basis is not the most relevant for solid state systems
such as perovskite oxides. For high symmetry situations, the eigenbasis
of the density matrix will be given by $t_{\text{2g}}$ ($xy,yz,xz$)
and $e_{\text{g}}$ ($3z^{2}-r^{2},x^{2}-y^{2})$ states. The conversion
matrix is
\[
\text{V}^{\,\sigma}=\left(\begin{array}{ccccc}
0 & 1/\sqrt{2} & i/\sqrt{2} & 0 & 0\\
0 & 0 & 0 & -i/\sqrt{2} & 1/\sqrt{2}\\
1 & 0 & 0 & 0 & 0\\
0 & 0 & 0 & i/\sqrt{2} & 1/\sqrt{2}\\
0 & 1/\sqrt{2} & -i/\sqrt{2} & 0 & 0
\end{array}\right)
\]
if we choose the order ($3z^{2}-r^{2},x^{2}-y^{2},xy,yz,xz$). The
transformed matrices are now in the more useful basis with entries
\begin{widetext}
\[
\text{A}^{\sigma}=\left(\begin{array}{r|rrrrr}
 & 3z^{2}-r^{2} & x^{2}-y^{2} & xy & yz & xz\\
\hline 3z^{2}-r^{2} & 0 & -0.517 & -0.517 & 0.517 & 0.517\\
x^{2}-y^{2} & -0.517 & 0 & 0.861 & -0.172 & -0.172\\
xy & -0.517 & 0.861 & 0 & -0.172 & -0.172\\
yz & 0.517 & -0.172 & -0.172 & 0 & -0.172\\
xz & 0.517 & -0.172 & -0.172 & -0.172 & 0
\end{array}\right),
\]

$ $and
\[
\text{B}^{\sigma}=\left(\begin{array}{r|rrrrr}
 & 3z^{2}-r^{2} & x^{2}-y^{2} & xy & yz & xz\\
\hline 3z^{2}-r^{2} & 1.143 & -0.630 & -0.630 & 0.059 & 0.059\\
x^{2}-y^{2} & -0.630 & 1.143 & 0.288 & -0.401 & -0.401\\
xy & -0.630 & 0.288 & 1.143 & -0.401 & -0.401\\
yz & 0.059 & -0.401 & -0.401 & 1.143 & -0.401\\
xz & 0.059 & -0.401 & -0.401 & -0.401 & 1.143
\end{array}\right).
\]

\[
\]
\end{widetext}

These matrices directly tell us how the $U|J$ scheme corrects the
energy eigenvalues beyond the $U_{\text{eff}}$ energy shift. For
example, the diagonal entries of $\text{B}^{\sigma}$ indicate that
occupying any orbital pushes up the energy of the opposite spin orbitals
by $1.14J$. 

As another example, if we have an ion such as Mn$^{4+}$ with a full
up spin and empty down spin $t_{\text{2g}}$ shell, so that $f_{\uparrow}=(0,0,1,1,1)$
and $f_{\downarrow}=0$, then for the up spin orbitals the energy
correction beyond $U_{\text{eff}}$ is $(0.52,0.52,-0.34,-0.34,-0.34)J$
which destabilizes the same spin $e_{\text{g}}$ and stabilizes the
same spin $t_{\text{2g}}$ while for spin down orbitals the situation
is exactly reversed with energy correction $(-0.52,-0.52,0.34,0.34,0.34)J$.
A final example is a full $t_{\text{2g}}$$^{6}$ shell such as Co$^{3+}$
which gives zero correction to the $U_{\text{eff}}$ scheme. The above
two matrices form the basis for various analyses in the main text.

\section*{Appendix B: Density matrix rotation to local axis representation\label{sec:Appendix-B}}

In typical DFT+\emph{U} approaches, the electronic structure is given
in terms of density matrices for each sub-space, \emph{e.g.}, the
\emph{d} shell. Unfortunately the orthogonal global axial representation
which is most efficacious for computation is often not most convenient
for analysis and understanding. This happens in calculations with
non-trivial unit cells where inequivalent oxygen octahedra surround
transition metal ions. Octahedral rotations and tilts mean the global
axial system for the calculation, here labelled $x',y',z'$, will
differ from the native local axes, labelled $x,y,z$. Native axes
for each octahedron point along the transition metal-O bonds, and
form the natural basis for understanding the electronic structure
of the transition metal \textit{d} orbitals. We describe the details
of a simple approach that rotates the density matrix, from the global
to the local basis \emph{via} polynomial transformations, with LaMnO\textsubscript{%
3%
} as our example.

We choose a particular Mn ion and its nearest neighbor O atoms which
identify an octahedral cage. Three Mn-O bonds are chosen that point
in approximately orthogonal directions. The bonds are indexed $i=1,2,3\,$,
and we compute the difference vectors from the Mn to O positions:
$\mathbf{u}_{i}=\mathbf{r}(O_{i})-\mathbf{r}(Mn)$. These vectors
are then normalized and define the local axes for the Mn. We create
a $3\times3$ rotation matrix $\text{R}$ connecting the global $x',y',z'$
and local $x,y,z$ systems 

\[
\left(\begin{array}{c}
x'\\
y'\\
z'
\end{array}\right)=\text{R}\left(\begin{array}{c}
x\\
y\\
z
\end{array}\right)=\left(\begin{array}{ccc}
R_{11} & R_{12} & R_{13}\\
R_{21} & R_{22} & R_{23}\\
R_{31} & R_{32} & R_{33}
\end{array}\right)\left(\begin{array}{c}
x\\
y\\
z
\end{array}\right)
\]
defined by placing the unit vectors $\mathbf{u_{i}}$ in the columns
of $\text{R}$. It is at this point that we choose the ordering of
the unit vectors to reflect the physical questions at hand\textit{.
}\textit{\emph{Note,}}\textit{ }a traditional choice is to align $z$
with the apical bond, but other choices are possible: for example,
in our work we have placed $y$ along the non-Jahn-Teller 'apical'
Mn-O (see Figure \ref{fig:Orbital-ordering-across-LMO}) while $x$
and $z$ span the Jahn-Teller active plane.

This rotation represents a linear polynomial transformation relating
$x',y',z'$ to $x,y,z$. The angular behavior of each \textit{d} orbital
is quadratic in the coordinates: $3z'^{2}-r'^{2},\, x'^{2}-y'^{2},\, x'y',\, y'z',\, x'z'$,
so it is straightforward to plug in and algebraically transform the
polynomials to the unprimed (local) coordinate system. Performing
the substitutions, using the orthogonal nature of the $\text{R}$
matrix, and collecting terms, we find

\[
\left(\begin{array}{c}
3z{}^{2}-r{}^{2}\\
x{}^{2}-y^{2}\\
xy\\
yz\\
xz
\end{array}\right)=\text{C}\left(\begin{array}{c}
3z'^{2}-r'^{2}\\
x'^{2}-y'^{2}\\
x'y'\\
y'z'\\
x'z'
\end{array}\right)
\]
where the matrix $\text{C}$ is 

\begin{widetext}

\begin{multline*}
\text{C}=\left(\begin{array}{ccccc}
\frac{1}{2}(3R_{33}^{2}-1) & \frac{1}{2}(R_{13}^{2}-R_{23}^{2}) & \frac{1}{2}R_{13}R_{23} & \frac{1}{2}R_{23}R_{33} & \frac{1}{2}R_{13}R_{33}\\
\frac{3}{2}(R_{31}^{2}-R_{32}^{2}) & \frac{1}{2}(R_{11}^{2}-R_{12}^{2}+R_{22}^{2}-R_{21}^{2}) & \frac{1}{2}(R_{11}R_{21}-R_{12}R_{22}) & \frac{1}{2}(R_{21}R_{31}-R_{22}R_{32}) & \frac{1}{2}(R_{11}R_{31}-R_{12}R_{32})\\
6R_{31}R_{32} & 2(R_{11}R_{12}-R_{21}R_{22}) & R_{11}R_{22}+R_{12}R_{21} & R_{21}R_{32}+R_{22}R_{31} & R_{11}R_{32}+R_{12}R_{31}\\
6R_{32}R_{33} & 2(R_{12}R_{13}-R_{22}R_{23}) & R_{12}R_{23}+R_{13}R_{22} & R_{22}R_{33}+R_{23}R_{32} & R_{12}R_{33}+R_{13}R_{32}\\
6R_{31}R_{33} & 2(R_{11}R_{13}-R_{21}R_{23}) & R_{11}R_{23}+R_{13}R_{21} & R_{21}R_{33}+R_{23}R_{31} & R_{11}R_{33}+R_{13}R_{31}
\end{array}\right).
\end{multline*}

\end{widetext}

The matrix $\text{C}$ is not unitary due to the fact that the bare
polynomials $3z'^{2}-r'^{2},x'^{2}-y'^{2},x'y',y'z',x'z'$ are orthogonal
but are not normalized. The normalization is done by averaging the
squares of the functions $(3z'^{2}-r'^{2})/r'^{2},(x'^{2}-y'^{2})/r'^{2},x'y'/r'^{2},y'z'/r'^{2},x'z'/r'^{2}$
over the surface of the unit sphere. We place these averages, which
are $4/5,4/15,1/15,1/15,1/15$, respectively, on the diagonals of
a diagonal scaling matrix $\text{S}$ and then form the scaled and
unitary transformation matrix $\text{D}=\text{S}^{1/2}\,\text{C}\,\text{S}^{-1/2}$
which is our final matrix relating the \textit{d} orbitals in primed
and unprimed coordinates.

To give a feeling for how the method works, we take the experimental
structure for LaMnO\textsubscript{%
3%
} crystal with $a=5.736\ {\rm \AA}$, $b=7.703\ {\rm \AA}$ and $c=5.540\ {\rm \AA}$,
as in Figure \ref{fig:exptlmostruct}. In experimental structured
LMO, consider the octahedron about the Mn atom at $(0.00,0.00,2.77)\ {\rm {\rm \AA}}$,
which has two basal oxygens at $O^{1}=(1.12,-0.31,1.26)\ {\rm {\rm \AA}}$
and $O^{2}=(1.75,0.31,4.03)\ {\rm {\rm \AA}}$, and an apical oxygen
at $O^{3}=(-0.07,1.93,2.37)\ {\rm \AA}$. We form the normalized $\mathbf{u}_{i}$
vectors, compute $\text{R}$ and then $\text{C}$ and upon normalization
find 

\[
\text{D}=\left(\begin{array}{ccccc}
0.06 & 0.04 & 0.01 & 0.23 & 0.31\\
-0.03 & 0.16 & 0.88 & -0.24 & 0.64\\
0.58 & -0.74 & 0.97 & -0.09 & 0.27\\
0.78 & 0.54 & -0.71 & -0.19 & -0.33\\
-0.05 & -0.20 & -2.76 & -0.29 & 0.99
\end{array}\right).
\]

$\text{D}$ can now be used to diagonalize the $5\times5$ density
matrix in the sub-space of the Mn \emph{d }orbitals. For a DFT+$U|J$
calculation with $U=8$ eV and $J=2$ eV, fixed at the experimental
structure, the Mn \emph{d} eigensystem is

\[
\phi_{\sigma}=\left(\begin{array}{r|rrrrr}
f_{i\sigma} & 0.42 & 0.96 & 0.97 & 0.98 & 0.99\\
\hline 3z'^{2}-r'^{2} & 0.50 & 0.12 & 0.80 & -0.22 & -0.19\\
x'^{2}-y'^{2} & 0.54 & 0.17 & -0.53 & -0.07 & -0.63\\
x'y' & -0.10 & 0.80 & -0.02 & -0.56 & -0.21\\
y'z' & 0.28 & -0.57 & 0.09 & 0.77 & -0.07\\
x'z' & -0.60 & 0.11 & 0.25 & 0.21 & -0.72
\end{array}\right).
\]
Here each eigenvector is a column vector with its eigenvalue $f_{i\sigma}$
provided above it. Before rotation, it is hard to easily read off
the nature of each eigenstate by inspection. After rotation, the eigenvectors
in the local basis are given by

\[
\text{D}\phi_{\sigma}=\left(\begin{array}{r|rrrrr}
f_{i\sigma} & 0.42 & 0.96 & 0.97 & 0.98 & 0.99\\
\hline 3z{}^{2}-r{}^{2} & -0.17 & -0.05 & -0.07 & 0.13 & -0.97\\
x{}^{2}-y{}^{2} & 0.99 & 0.05 & -0.07 & 0.03 & -0.17\\
xy & 0.00 & -0.01 & 0.05 & -0.99 & -0.15\\
yz & -0.03 & -0.99 & -0.09 & 0.00 & 0.07\\
xz & 0.01 & +0.11 & -0.99 & -0.06 & 0.07
\end{array}\right).
\]

The local basis eigenvectors are clearly much ``purer'' as each
vector has a component whose magnitude is $0.97$ or larger. And thus
each configuration is easy to read off by inspection: the partially
occupied state in the first column is essentially the $x{}^{2}-y{}^{2}$
state while the last column shows that the $3z{}^{2}-r{}^{2}$ has
become filled. We have strong orbital polarization in the $e_{\text{g}}$
manifold. This indicates the rotation to local octahedral coordinates
successfully diagonalized the eigensystem, and that the local basis
is physically relevant for understanding the electronic structure.%

\end{document}